\PassOptionsToPackage{usenames,dvipsnames}{color}
\PassOptionsToPackage{table}{xcolor}
\PassOptionsToPackage{hyphens}{url}
\documentclass[letterpaper,twocolumn]{sig-alternate-10pt}

\usepackage[numbers]{natbib}
\usepackage{epsfig,endnotes,epstopdf}
\usepackage{makecell}
\usepackage{amsfonts,xspace}
\usepackage[english,plain]{fancyref}
\usepackage{times}
\usepackage{supertabular}
\usepackage{rotating}
\usepackage{subcaption}
\usepackage{multirow}
\usepackage{graphicx}
\usepackage{tabulary}
\usepackage{comment} 
\usepackage{multirow,pifont,wasysym,fancyhdr,balance,capt-of} %dblfloatfix
\usepackage{hyperref}
\usepackage[T1]{fontenc}
\usepackage{tabularx,xspace}
\usepackage{pgfplots}
\usepackage{colortbl,booktabs}
\usepackage[titletoc,title]{appendix}

\newcommand{\ie}{{\it i.e.,}\xspace}
\newcommand{\eg}{{\it e.g.,}\xspace}
\newcommand{\etc}{{\it etc.}}
\newcommand{\etal}{\emph{et~al.}}
\newcommand{\geolocation}{{geolocation}\xspace}
\newcommand{\sysname}{{\textsc{Passport}}\xspace}
\newcommand{\ensemble}{{\it ensemble}\xspace}

\newcommand{\eat}[1]{}
\newcommand{\para}[1]{{\vspace{2pt} \bf \noindent #1 \hspace{10pt}}}

\newcounter{packednmbr}

\newcommand{\postfigspace}{-1.5em}

\newcommand{\numTrainSize}{11,626\xspace}
\newcommand{\numTrainCountrySize}{120\xspace}
\newcommand{\pctPredSingleCountry}{$88.1$\%\xspace}
\newcommand{\pctPredTwoCountry}{$95.5$\%\xspace}
\newcommand{\alidadeSourcesMonthYear}{June 2016\xspace}
\newcommand{\numClassifiers}{$4$\xspace}
\newcommand{\numTrainAS}{1,244\xspace}

\newcommand{\tabletextsize}{\footnotesize}
 
\newcommand{\drc}[1]{{\color{red}[DRC: #1]}}
\newcommand{\muzammil}[1]{{\color{blue}[MUZ: #1]}}

\newcommand{\todo}[1]{{\color{green}[MUZ: TODO: #1]}}
\newcommand{\muz}[1]{}

\newenvironment{packeditemize}{\begin{list}{$\bullet$}{\setlength{\itemsep}{0.2pt}\addtolength{\labelwidth}{-4pt}\setlength{\leftmargin}{\labelwidth}\setlength{\listparindent}{\parindent}\setlength{\parsep}{1pt}\setlength{\topsep}{0pt}}}{\end{list}}

\pdfpagewidth=\paperwidth
\pdfpageheight=\paperheight

\begin{document}
%\title{\sysname: Enabling Accurate Country-Level Router Geolocation using Inaccurate Sources} 
%\title[\sysname]{\sysname: Enabling Accurate Country-Level Router Geolocation using Inaccurate Sources} 
%\title{\sysname} 
%\subtitle{Enabling Accurate Country-Level Router Geolocation using Inaccurate Sources}
%\titlenote{Produces the permission block, and copyright information}
%\subtitle{Extended Abstract}
%
\title{{\fontsize{20}{20}\selectfont\textbf{P}}{\fontsize{16}{24}\selectfont\textbf{ASSPORT}}: Enabling Accurate Country-Level Router Geolocation using Inaccurate Sources} 

%\numberofauthors{1}
%\author{Paper \# 46, 14 pages}
\numberofauthors{3} 

\author{
	\alignauthor
		Muzammil Abdul Rehman\\
		Northeastern University\\
		muzammil@ccs.neu.edu
% 2nd. author
	\alignauthor
		Sharon Goldberg\\
		Boston University\\
		goldbe@cs.bu.edu
% 3rd. author
	\alignauthor
		David Choffnes\\
		Northeastern University\\
		choffnes@ccs.neu.edu
}

% \author{Firstname Lastname}
% \authornote{Note}
% \affiliation{%
%   \institution{Affiliation}
%   \streetaddress{Address}
%   \city{City} 
%   \state{State} 
%   \postcode{Zipcode}
% }
% \email{email@domain.com}

% \author{Firstname Lastname}
% \affiliation{%
%   \institution{Affiliation}
%   \streetaddress{Address}
%   \city{City} 
%   \state{State} 
%   \postcode{Zipcode}
% }
% \email{email@domain.com}

% \author{Firstname Lastname}
% \affiliation{%
%   \institution{Affiliation}
% }
% \email{email@domain.com}

% \author{Firstname Lastname}
% \affiliation{%
%   \institution{Affiliation}
% }
% \email{email@domain.com}

% \author{Firstname Lastname}
% \affiliation{%
%   \institution{Affiliation}
% }
% \email{email@domain.com}

%\renewcommand{\shortauthors}{F. Lastname et al.}
%\renewcommand{\shortauthors}{Paper \# 174}

\maketitle

% Use the following at camera-ready time to suppress page numbers.
% Comment it out when you first submit the paper for review.
%\thispagestyle{empty}

%\sloppy
\subsection*{Abstract} 
\label{sec:abstract}
%\begin{abstract}

When does Internet traffic cross international borders?  This question has major geopolitical, legal and social implications and is surprisingly difficult to answer.  A critical stumbling block is a dearth of tools that accurately map \emph{routers} traversed by Internet traffic to the \emph{countries} in which they are located.  This paper presents \sysname: a new approach for efficient, accurate country-level router geolocation and a system that implements it.  \sysname provides location predictions with limited active measurements, using machine learning to combine information from IP geolocation databases, router hostnames, whois records, and ping measurements. We show that \sysname substantially outperforms existing techniques, and identify cases where paths traverse countries with implications for security, privacy, and performance.

\section{Introduction}
\label{sec:introduction}

Determining when Internet traffic crosses international borders is of significant interest to both lawmakers and the general public. A nation's laws usually apply even to foreign traffic that transits that nation.  The US, UK, and other nations aggressively surveil traffic that crosses their national borders~\cite{FISAs702,ghcqFiber,germanFiber}.
Foreign traffic that traverses countries with aggressive censorship policies can be filtered before it arrives at its destination~\cite{anon2012collateral}.
Countries have debated (\eg Brazil~\cite{brazil2014loc}), 
%cite china, https://www.bloomberg.com/news/articles/2016-01-21/a-cybersecurity-law-in-china-squeezes-foreign-tech-companies
%
or enacted (\eg China~\cite{china2017loc}, Russia~\cite{Guardian:2015:russia-surveillance})
``data residency'' laws that require their citizens' data to remain on domestic soil.
In Europe, the Data Protection Directive~\cite{EU:1995:dataprotection,EU:2016:dataprotection,EU:transfers} forbids the movement of citizens' data to countries that do not provide ``adequate'' data protection.
In the US, the laws relating to the government's ability to search and seize their own citizens' data depend on whether that data was intercepted on US soil or abroad~\cite{Daskal:2015:territoriality}.

Despite increasing interest from the public, networking researchers still lack the tools needed to accurately determine where Internet traffic crosses international borders.
%Despite increasing interest from the public, researchers still lack the tools needed to accurately determine where Internet traffic crosses international borders.
%
While traceroute trivially reveals the IP addresses of the routers traversed by network traffic,
geolocating these \emph{router} IPs to   \emph{countries} remains a key open problem.
There are a number of commercial databases~\cite{ipinfo,maxmind,dbip,eurekapi,ip2location,google-geoloc,edgescape}
and research projects~\cite{wang2011towards,padmanabhan2001investigation,wong:octant,ethankb:geoloc, laki:spotter}
that accurately geolocate IP addresses belonging to \emph{eyeball} networks, Internet exchange points (IXPs),
and large enterprise datacenters. However, the accuracy of these sources on \emph{routers} is poor~\cite{huffaker:infrastructure-geolocation, poese:ip-geolocation-unreliable}. We find that when mapping routers to countries, even the best geolocation databases achieve more than 90\% accuracy for only
46.5\% of the countries traversed by paths in our traceroute dataset. Worse, na{\"i}vely mapping router IPs to countries using BGP paths and AS-level geolocation, has this accuracy for only 5\% of countries. Research projects that rely on active measurements can 
require large numbers of probes to be accurate, and their online deployments either fail to resolve many router locations~\cite{laki:spotter, wong:octant}, or is not unavailable.

To address this problem, we present \sysname, a system that accurately maps
router IPs to countries. \sysname identifies the set of countries traversed by a traceroute
path by combining small numbers of round-trip time (RTT) latencies with a machine learning classifier that uses IP geolocation databases.
\sysname first trains a machine-learning classifier on a set
of (individually inaccurate) geolocation sources, and uses the classifier to map the IP in a path to a set of countries.
These results are filtered through speed-of-light (SoL) constraints imposed by RTT latencies.
This enables  \sysname to filter out incorrect predictions %$[HOW DO WE KNOW THIS? WHAT IS OUR GROUND TRUTH?] 
and maps each IP
 to a single country \pctPredSingleCountry of the time. 
\muz{case studies}

Our main contributions are as follows.
\begin{packeditemize}
\item We build a dataset of \numTrainSize ground-truth router IP geolocations from \numTrainAS autonomous systems (ASes) in  \numTrainCountrySize countries, and use them to evaluate the accuracy of existing approaches for router geolocation. These locations come from well-established data sources such as IXP locations, crowdsourced labels from operators, and reverse DNS entries, and they are all cross-validated. % using speed-of-light constraints.
\item We design and evaluate an empirically informed approach to aggregating a suite of machine-learning classifiers for predicting a router's country, using our ground truth labels for training. More specifically, we demonstrate that it is possible to achieve high country-level geolocation accuracy from individually inaccurate data sources by combining independent classifiers trained on different subsets of a possibly-biased dataset.
\item We build and evaluate techniques that incorporate active measurements and iterative learning that improve classifier precision and accuracy.
\item We use our system to analyze the geopolitical properties of targeted intra-country and international paths, identifying interesting cases and their implications. For example, BRICS (Brazil, Russia, India, China, and South Africa) nations, among others, often wish to avoid transiting traffic through the US, where it can be subject to surveillance. However, we find that paths between Brazil and Russia, as well as those between China and India, transit the US. Russia, which reportedly meddled in recent European elections~\cite{marconRussia,germanyElectionRussia}, transits traffic for several paths between European countries. Even ``purely domestic'' traffic detours outside of country where sources and destinations are located, including a case where domestic Philippines traffic transits Hong Kong. 
\item In addition to providing an online tool to provide country-level IP geolocation, we will make all of our code and data publicly available.
\end{packeditemize}

% Muz: Cutting 
\begin{comment}

\drc{cut?}In the next section, we discuss prior work on geolocating Internet paths (\S\ref{sec:related}),
then we detail the  goals, assumptions, and design of \sysname  (\S\ref{sec:goals}).
We present our implementation (\S\ref{sec:implementation}) and evaluate its geolocation accuracy (\S\ref{sec:evaluation}).  We use \sysname to identify Internet paths with interesting geopolitical properties  (\S\ref{sec:analysis}),
and conclude with a discussion of open problems and future work (\S\ref{sec:discussion}). 

\end{comment}

\section{Related Work and Motivation}
\label{sec:related}

This section discusses prior IP geolocation work. We exclude the use of GPS, which is now commonly accessible when measuring from mobile devices, because such information is generally \emph{not} available for router IPs.
	
\para{Constraint-based schemes.} These schemes~\cite{wang2011towards,padmanabhan2001investigation,wong:octant,ethankb:geoloc} primarily use speed of light (SoL) constraints based on RTT measurements from given landmark locations to identify regions where a router can feasibly be located. Each approach uses some source of ground truth to further tighten the contraints on where a host may be geolocated. 
 Eriksson~\etal ~\cite{eriksson:learninglocation} use RTTs and a Naive Bayesian classifier to identify router cities and counties (not \textit{country}), but use an inaccurate geolocation database~\cite{maxmind} to validate their results. Posit \cite{eriksson:posit} uses statistical analysis on latency measurements and landmarks to identify possible locations.

\para{Hostname Parsing schemes.} The hostnames associated with router IPs often encode location information, \eg \url{ae-4-90.edge5.frankfurt1.level3.net} encodes the city of Frankfurt, Germany. Previous work leverages this to provide mappings from router hostnames to their locations via simple matching~\cite{undns} and machine learning~\cite{drop}. These approaches generally provide accurate IP geolocation at the regional level, but can be inaccurate when there is location ambiguity (\eg { \tt san} can indicate the airport code for San Diego, California, or the start of several cities starting with ``San'' such as San Juan). Further, this technique only works if location-encoded hostname information is available for the IP address. The IXMap project~\cite{ixmaps} uses a combination of hostname parsing and next/previous hop RTT latency to assign router locations.
%Google, for instance, does not provide any hostname information and hence this technique fails for Google's IP addresses.

\para{Geolocation Databases.}
Several free and paid services offer databases that map IP addresses to locations~\cite{ipinfo,maxmind,dbip,eurekapi,ip2location,google-geoloc}. 
%While previous work indicates they are generally consistent in predicted locations~\cite{Huffaker:2011}, 
However, recent work indicates that they have limited accuracy for geolocating infrastructure IP addresses~\cite{huffaker:infrastructure-geolocation}. In part to address this, OpenIPMap~\cite{openipmap} maintains a crowdsourced list of router IP geolocations, but is limited to covering IPs provided by contributors. 
%Recent work uses layer-1 physical topology maps~\cite{durairajan:layer1} to better understand how to exhaustively measure ISP infrastructure.
%
Alidade~\cite{alidade} uses a collection of databases and measurements to identify router locations. In contrast, we develop an adaptive strategy for incorporating unreliable information to provide reliable predictions of the \textit{country} where a router IP is located. 
%In contrast to this work, we develop a comprehensive, adaptive strategy for incorporating unreliable location database information to provide reliable  predictions of the \textit{country} where a router IP is located. 

%\para{Network positioning.} Bunch of stuff from my prior work~\cite{ajsu:cdn-rel-pos,dabek:vivaldi,francis:idmap,wong:meridian}.

\para{Geopolitical routing implications.}  When Internet traffic traverses national borders, it may be subject to surveillance~\cite{eo12333,Daskal:2015:territoriality,germanFiber,ghcqFiber,FISAs702} and censorship regimes~\cite{anon2012collateral}. As a result, the privacy and integrity of users' Internet traffic depends not only on endpoint locations, but also on the location of intermediate hops. In addition, large geographical detours can turn into path inflation that substantially impacts end-to-end performance~\cite{whyhigh}.

\para{Summary.} Previous approaches rely on independent, complementary approaches to geolocating IP addresses, none of which alone has sufficient accuracy to reliably geolocate the country where a router IP resides. As a result, researchers are currently in the dark regarding the important question of which countries Internet traffic traverses. In our work, we leverage the observation that each of these approaches has different strengths and weaknesses, so there is an opportunity to combine them to provide greater accuracy as a whole than any individual part. The next section describes how we leverage this observation by using machine learning principles to reliably predict router-IP countries.

\section{Goals and Design}
\label{sec:goals}

\begin{figure}[t]
    \centering
    \begin{subfigure}[t]{0.9\linewidth}
        \includegraphics[width=1.0\linewidth]{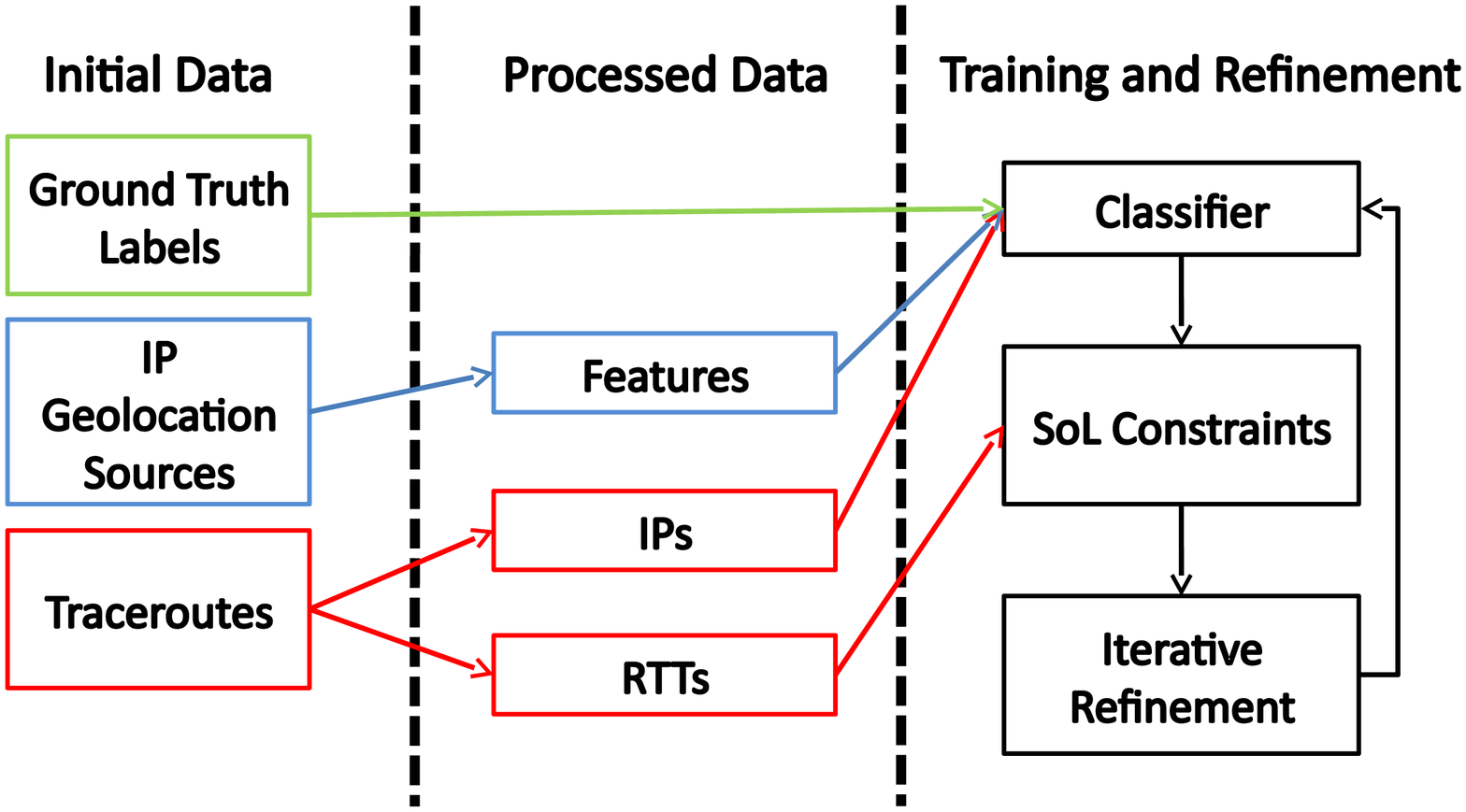}
        \vspace{-2em}
        \caption{Offline training.}
        \label{fig:offline_stage_design}
    \end{subfigure}

    \begin{subfigure}[t]{0.9\linewidth}
        \includegraphics[width=1.0\linewidth]{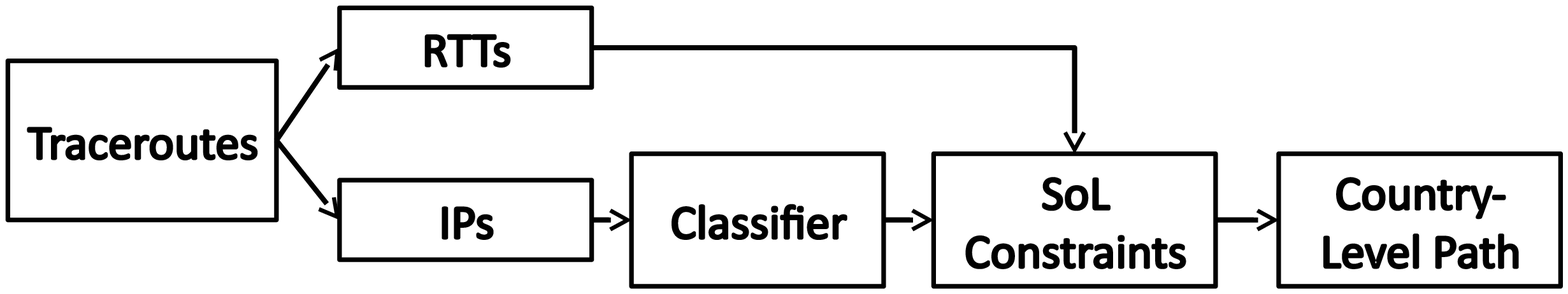}
        \vspace{-2.2em}
        \caption{Online prediction.}
        \label{fig:online_stage_design}
    \end{subfigure}
    \caption{ \textbf{Key system components.} {\sl The offline training module (top) builds a machine-learning classifier that  predicts router country; the online prediction module (bottom) provides interactive prediction of router countries for an input set of traceroute measurements.}}
    \vspace{\postfigspace}
    \label{fig:diagram-overview}
\end{figure}

\subsection{Goals}

Our primary goal is to build a system that can \textit{accurately identify the countries of Internet routers that respond to active measurements probes (e.g., pings).}
Our ancillary goals are to provide an online system that makes predictions quickly so that it can map both transient and long-lived paths, and to ensure that the system quickly adapts to changes in Internet topologies that affect router-location predictions. 
The system should be accurate enough to understand the \textit{geopolitical} properties of Internet paths regarding security and privacy.

% that can adapt to dynamic changes in the internet topology, provided a set of inaccurate \geolocation sources and traceroute measurements to these IP addresses}. 
%Since IP addresses can be reassigned to new routers, therefore, the current location of these IP addresses should be predicted over time and this system should be able to adaptable to changes in the inaccurate data sources. 
%The mapping of routers should also be done in \textit{real-time} or atleast \textit{near real-time} as IPs may be reassigned to newer locations. 
%We aim to study the \textit{geopolitical} nature of the internet and the possible consequences of paths that internet traffic might through different geographical boundaries take as it goes from a source to a destination.

%While geolocating so-called ``eyeball'' IP addresses is well studied problem with a variety of databases and tools for providing high-precision location information,  
Geolocations of \emph{router IPs} have received little attention compared to ``eyeball'' IPs;  
in fact, na{\"i}ve approaches to using geolocation databases for such infrastructure IPs leads to high inaccuracy (\eg 29\% of IP geolocations are incorrect in our labeled dataset for MaxMind, shown in Table~\ref{tab:classifier_features}). When looking at the accuracy on a per-country level, existing approaches achieve 90\% accuracy or better only for 5--46.5\% of countries (Figure~\ref{fig:results_classifier_comparison_accuracy_ensemble}). Such inaccuracy can significantly impact our ability to correctly interpret geopolitical implications of paths traversing those routers.
 
%\drc{Incorporate the text below in a much more specific way.}
% the \geolocation sources to predict these locations are relatively inaccurate, as shown in Figure \ref{fig:results_classifier_comparison_accuracy_ensemble}. 
%Some of these \geolocation sources have a lower accuracy than others.

Thus, instead of relying on any fixed set of geolocation data sources to predict a router location, we take an alternative approach that relies on machine learning. 
Specifically, our hypothesis is that the set of geolocation sources that will reliably predict a router's country varies according to properties of the router (\eg IP, BGP prefix, \etc), and that we can build a machine-learning classifier that reliably predicts router locations based on an ensemble of individually unreliable sources. 
We test this hypothesis and show that a machine-learning classifier, combined with active measurement probes, can achieve substantially higher accuracy and precision compared to previous approaches (\eg at least 90\% accuracy for 96.5\% of countries, as shown in Fig.~\ref{fig:results_classifier_comparison_accuracy_ensemble}).  

%Mapping the locations of the routers at the \textit{core} of the Internet\drc{capitalize Internet everywhere, always} to countries is an inherently difficult problem. Although, end-user locations can be accurately measured\muzammil{refer to papers from section 2}, however, finding router locations is relatively difficult. 
%As discussed in Section \ref{sec:related}, router locations are kept private by the ISPs in general. 
%This information is not publicly available and the \geolocation sources to predict these locations are relatively inaccurate, as shown in Figure \ref{fig:results_classifier_comparison_accuracy_ensemble}. 
%Some of these \geolocation sources have a lower accuracy than others.

\para{Assumptions.}
Our system takes as input a set of traceroute measurements (along with round-trip times to each responsive router) and a collection of IP geolocation data sources. After processing this initial data, we assign countries to each IP address\footnote{Hereafter referred to simply as \emph{IP}.} along a path that we observe. We assume that the errors affecting each data source are not random and that data sources achieve high accuracy for at least some networks. We further assume that our ground truth IP geolocations are correct. Though we focus on country-level geolocation, our input geolocation sources can use finer-grained precision.
 
To bootstrap our traceroute-based country-level geo\-location analysis, we assume that fixed-line end-host geolocations can be predicted accurately. We also assume that each IP observed along a path is assigned to exactly one corresponding router,\footnote{As such, we will use the terms \emph{router} and \emph{IP} interchangeably.} and that neither the router nor the IP change locations substantially during each measurement and analysis round, which is currently one day. While it is certainly the case that IPs can be reassigned or reused arbitrarily within an ISP, we expect it to rarely affect our conclusions.

%and this information can be used to choose a source to issue traceroute measurements as well as being used as a \textit{landmark} (a known location that can be used to measure distances on a map).

% keep for reference: \drc{NEVER use contractions in a paper. Also, ``the paper'' doesn't do anything because it's a paper. Never use it as a subject. Instead use ``Our approach'' or ``This work'' or ``We do not''. Always add a space befre "(". never use connotations like "don't"}

% 

\para{Non-goals.} This work does not focus on geolocating eyeball IP addresses; rather, we assume that geolocation databases provide high accuracy for such IPs and thus use them as ``anchors'' for geolocating routers. 
We do not map routers to city-level (or finer-grained) locations; rather we focus only on the country where it resides. 
This is sufficient to inform several important security and privacy analyses.
We do not attempt to provide perfect accuracy or coverage of Internet paths; however, our approach should geolocate most Internet paths most of the time.
If a router on a path does not respond to a traceroute (or a ping) probe, then we cannot use SoL to geolocate it; however, we can use SoL constraints to other responding hops on the path to identify the set of countries such absent routers might be located in. 

%We aim to map router locations to a country-level using a set of inaccurate \geolocation sources and use this information to look at the geographical boundaries traversed. <---- This is a goal, not a non-goal. Not sure why you put it here.  
%This is our primary difference from past research\muzammil{refer to papers from section 2}

%\artbd{Note one line contains one sentence in this tex file. This is useful to track diffs between successive commits.}

\begin{table*}[!th]
\setlength\tabcolsep{2pt}
\tabletextsize
	\centering
	\rowcolors{2}{gray!10}{white}
		\begin{tabular}{l | l | c | p{4in} }
			\textbf{\S} & \textbf{Topic} & \textbf{Dataset} & \textbf{Key results} \tabularnewline \hline
			%\ref{subsec:offline-training}
			\ref{subsubsec:methodology-classifier} & Classifier Selection & Controlled exp. & We develop an \ensemble of \numClassifiers classifiers, trained on different subsets of labeled data. \tabularnewline			
			\ref{subsec:overall-accuracy} & Overall Accuracy & Ground truth & We show that our  approach is more accurate than single classifiers and geolocation services.  \tabularnewline
			\ref{subsubsec:ensemble-construction} & Ensemble Construction & Ground truth & Adding more classifiers in the \ensemble provides diminishing returns for the increase in accuracy while increasing the number of countries predicted (decreasing precision). \tabularnewline		
			\ref{subsec:constraint-refinement} & Constraint-based Refinement & Traceroute IPs & We increase the precision of the \ensemble classifier in \sysname using SoL constraints (\sysname maps \pctPredSingleCountry IPs to a single country and \pctPredTwoCountry to at most two countries). \tabularnewline
			\ref{subsubsec:results-comparision-alternate} & Comparison: Alternatives & Traceroute IPs & \sysname has a high consistency with EdgeScape and IP2Location, but lower with other geolocation databases. Most inconsistencies occur when geolocation databases predict a country that violates SoL constraints, and these affect accuracy for large fractions of paths.  \tabularnewline
			\ref{sec:analysis} & Geopolitical Case Studies & Traceroute IPs & \sysname identifies many cases of international detours, reverse-forward path asymmetry, and circuitous paths. \tabularnewline
		\end{tabular}
		\caption{\textbf{Roadmap for key topics covered in \S\ref{sec:implementation}, \S\ref{sec:evaluation} and \S\ref{sec:analysis}.}}
		\label{tab:roadmap}
\vspace{\postfigspace}

%\vspace{-2.5em}
\end{table*}

%We now discuss the key aspects of our design and implementation. We then
%evaluate our design choices and use \sysname on large-scale traceroute measurements for case studies involving international and transcontinental paths. 

\subsection{Design}
\label{sec:design}

\sysname is designed around two high-level components, \emph{offline training} and \emph{online prediction} (Figure~\ref{fig:diagram-overview}). For offline training, we begin with traceroutes, ground truth location labels, and IP geolocation databases. We use these as input to train a machine-learning classifier that predicts the country where a router is located. 

For online prediction, we accept as input (from a user, or via an API call) a traceroute measurement and return the set of countries in which each router is predicted to be located. The result can be zero or more countries, but as we show in \S~\ref{sec:evaluation}, we predict exactly one country for the vast majority (88.1\%) of cases. This section provides a high-level overview of the system design; we cover implementation details in \S~\ref{sec:implementation}. Table~\ref{tab:roadmap} presents a roadmap for the remainder of the paper.

\subsubsection{Offline Training}
The purpose of the offline training component of \sysname is to build a machine-learning classifier that reliably predicts the country-level location of routers appearing in arbitrary traceroute measurements. Offline training consists of three phases (Fig.~\ref{fig:offline_stage_design}): data collection, feature selection, and training and refinement. Data collection entails gathering router IPs and RTTs from traceroutes, geolocation hints from available (unreliable) sources, and ground truth geolocations when available. 
%As the set of available geolocation hints can be large\drc{how big?}, we perform feature selection\drc{do we? not mentioned later} to identify the features that are most useful for reliable router-country predictions. 

In the training and refinement phase, we first use ground truth labels (\S\ref{sec:dataset}) and features to train an initial classifier. We then use RTTs from traceroute data to rule out any classifier predictions that violate speed of light (SoL) constraints. As a result, there may be cases where at least one router has no predicted country. In the iterative refinement phase, we use the geolocated routers on a path and SoL constraints to help locate such routers. We then use the results of these analyses to \emph{retrain} the classifier. This phase terminates when the set of predicted router locations becomes stable (\eg no more than $1$\% of router locations change from one iteration to the next). In practice our system converges within two or three rounds. We run this analysis periodically, currently using a (configurable) period of one week. 

Note that our design incorporates a single logical classifier, but can be (and is) composed of an ensemble of classifiers. 
We refer to this single logical classifier as \ensemble. 
As we discuss in the next section, we use a custom ensemble of classifiers, with each sub-classifier selected based on empirical analysis to achieve high accuracy (see Appendix \ref{sec:appendix:ensemble}, to make router-country predictions. 

\subsubsection{Online Prediction}
The online prediction component makes router-country predictions interactively (Fig.~\ref{fig:online_stage_design}). The input to this component is a set of traceroutes to be geolocated at the country level. We use the classifier trained in the offline phase to make initial predictions for router locations. Much like the refinement phase in the offline training component, we impose SoL constraints to identify feasible country locations for routers. The output of online prediction is a traceroute path labeled with router countries.  See Appendix \ref{sec:appendix:online-system-website} for an example.

\section{Implementation}
\label{sec:implementation}

We now discuss the implementation of each component of the \sysname design.

\subsection{Offline Training}
\label{subsec:offline-training}
The offline training component currently runs once per week, a tunable parameter. It takes as input both ground truth and unreliable location data, extracts features for prediction, and uses them to build a machine-learning classifier. We now discuss the implementation details. 

\subsubsection{Data Sources}
\label{sec:dataset}

%Our data sources can be categorized in two main: ground truth and \geolocation features. 
%The ground truth dataset are the data sources that we use as \textit{labels} to train our classifier, while the \geolocation features are classifier \textit{features} available for most hosts.
We use the data below to train classifiers in \sysname. We identify the location of \numTrainSize IP addresses, in \numTrainCountrySize countries and \numTrainAS ASes (Table \ref{tab:ground_truth}).

\begin{table}[t]
\tabletextsize
	\centering
	%\rowcolors{2}{gray!10}{white}
	\begin{tabular}{ l | r | r | r}
		%\hline
		\textbf{Source} & \textbf{IPs} & \textbf{Countries} &\textbf{ASes} \\ 			\hline
		\emph{Union of all sources}                 & \emph{\numTrainSize}    & \emph{\numTrainCountrySize}      & \emph{\numTrainAS}       \\
		\ \ \ OpenIPMap                            & 8,973     & 117   &   1,037      \\ 
		\ \ \ Manually Labeled                    & 2,422    & 58    & 668        \\
		\ \ \ IXP                                  & 231          & 11   & 7    \\
		% \hline
	\end{tabular}
	\caption{\textbf{Ground truth IP geolocations.} {\sl Our dataset covers a substantial number of IPs worldwide.}}
	\vspace{\postfigspace}
	\label{tab:ground_truth}
\end{table}

\para{Reliable location data.} We assume the following datasets to be reliable and use them as ground truth. First, we use the lists of IXP addresses and locations provided by Packet Clearing House~\cite{pch:ixp}. Such prefixes and locations are well known and tend to change rarely. 

Second, we use crowdsourced labels from  OpenIPMap~\cite{openipmap} gathered between May 2016 and March 2017, comprising 19,257 IP addresses.
OpenIPMap is a publicly available database that uses crowdsourcing to geolocate IP addresses, and as such may contain incorrect labels. We used RTT measurements from April, 2017, to throw out any infeasible IP geolocations  from OpenIPMap as well as those that were mapped to a region spanning multiple countries. After this filtering step, we are left with 8,973 addresses. We cross-validated the OpenIPMap dataset and found that using it improved our coverage (by 62 countries) and country-level accuracy (nearly doubling the number of countries where we have 80\% or higher accuracy) even if a significant number of labels are incorrect. In fact, our analysis showed that even when 10\% or 20\% of OpenIPMap locations are incorrect, the accuracy of \sysname only decrease by 0.6\% and 4.1\%, respectively. The details of this analysis are available in Appendix \ref{sec:appendix:open-ip-map-eval}.

\para{Selecting IPs for manual labeling.}
To seed our system with useful measurement data spanning a diverse set of commonly traversed countries and ISPs, we conducted an initial set of traceroute measurements for the purpose of manually labeling routers with their locations. These measurements were conducted in February, 2017.  We selected 67 PlanetLab vantage points as sources, covering 20 countries. For each source, we picked the corresponding country-specific Alexa Top-100 websites as destinations. For each traceroute, we removed the endpoints and identified the countries where each intermediate hop IP was registered according to \emph{whois}. 

This set of 13,744 IPs covered a large set of countries around the world, but there were too many to manually label. We thus sampled 
the IPs using a set of heuristics to cover IPs in distinct prefixes across a large number of ASes globally. Specifically,  we identified the top 20 largest ASes\footnote{In countries with 20 or fewer ASes registered, we selected all of the ASes.} in each country (in terms of the customer cone size based on CAIDA AS Rank\cite{asrank} data), then picked a destination IP address from up to five routable prefixes in the AS.\footnote{For ASes with fewer than five prefixes, we selected one IP  for every prefix.} We did this for the 50 countries with the largest number of \emph{whois}-registered IPs, yielding 2,653 IPv4 addresses. Of these, 231 belonged to IXPs with public locations~\cite{pch:ixp}, leaving 2,422 unlabeled routers (from 668 ASes) for manual labeling.

Note that this approach was heuristically designed to cover a set of diverse networks worldwide with a limited amount of probes, and we do not claim to have maximized coverage or diversity. Rather, the point of this dataset is to provide useful initial data for training our classifier. Approaches that improve the scale and diversity of labeled routers should only further improve our system.

\begin{figure}[tb]
	\centering
	\includegraphics[width=0.99\columnwidth]{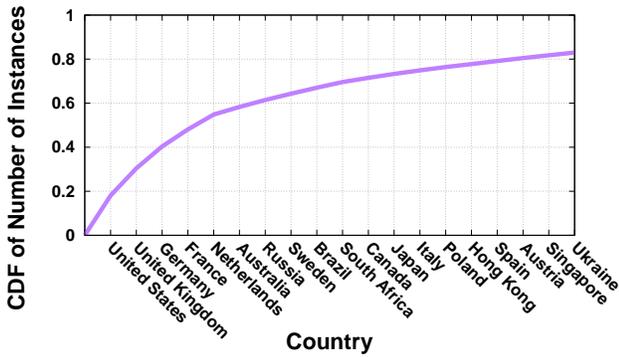}
	\caption{\textbf{Top 20 countries in terms of \# of routers in the ground truth dataset}. {\sl There is bias toward the US and European countries, which we need to account for when training classifiers.}} %
	\vspace{\postfigspace}
	\label{fig:ground_truth_countries_stats}
\end{figure}

\para{Ground truth labels.}
We manually labeled the location of 2,422 IP addresses. %in the above traceroute dataset in two phases. 
%First, we label the locations of \emph{endpoints} (\ie hosts that are not routers) using the two IP geolocation databases with the highest overall accuracy in our experiments (IP2Location and IPInfo), because such databases are generally accurate at locating end hosts. Next, we use manual analysis to validate those locations. 
%
First, we identify the country for each unlabeled router using targeted traceroutes toward the router from multiple vantage points (first from a set located in different continents, then if necessary using ones in the same continent as the unlabeled router), and incorporate geolocation data inferred from corresponding RTT-based SoL constraints and router hostnames. For routers with hostnames that encode geolocation information (\eg via city names or airport codes) we use both DRoP~\cite{drop} and our own manual analysis to identify each router's country.  

Figure~\ref{fig:ground_truth_countries_stats} shows a CDF of the fraction of IPs covered by the top 20 countries\footnote{The complete list of countries in terms of the number of routers in the ground truth is available at \url{https://goo.gl/umsbjz}}. Interestingly, the top 3.3\% (4 out of 120) countries cover 48.1\% of the IPs. We discuss techniques to account for this bias below.

%Figure~\ref{fig:ground_truth_countries_stats} shows a CDF of the fraction of IPs covered by the top 20 countries (see the full list in Appendix \ref{sec:appendix:ground-truth}). Interestingly, the top four countries (out of \numTrainCountrySize) cover 48.1\% of the IPs. We discuss techniques to account for this bias below.

%\muz{added} \para{Limitations of ground truth.}  Our ground truth dataset covers a total of \numTrainSize IP addresses in \numTrainCountrySize countries. This dataset is significantly diverse in the number of countries, but still can be considered small. The problem of labeling IP addresses to exact countries is relatively terse and less rewarding. Although, the dataset biased in favor of a few countries, we develop the \ensemble classifier in \sysname such as to minimize the effect of such biases and predict a country despite these setbacks. \muz{ends}

\para{Unreliable geolocation sources.}  Our input dataset includes several unreliable geolocation sources: IP geolocation services (IPInfo~\cite{ipinfo}, Maxmind Geolite2~\cite{maxmind}, DB-IP~\cite{dbip}, APIgurus~\cite{eurekapi}, IP2Location~\cite{ip2location}), hostname parsers (DDec~\cite{ddec}, which combines DRoP~\cite{drop} and undns~\cite{undns}) and AS information (\emph{whois} entries and AS Rank~\cite{asrank}). 

While geolocation services have been used as reliable sources for end host IPs, our analysis in \S\ref{sec:evaluation} shows that they provide low accuracy for router IPs in our dataset. Hostname parsers can have high accuracy for routers, but they also offer low coverage and have false positives when hostnames are ambiguous. \emph{whois} data contains the country where an AS or IP prefix is registered, but that often differs from where a corresponding router is located. Next, we describe how we extract features from these unreliable sources and use them as input to a machine-learning classifier.

%For manual analysis, we manually performed traceroutes to the IP addresses and decided the locations based upon next-hop routers, RTT and the hostnames of the routers.\drc{This is still not convincing.} \muz{Hostnames generally contain hidden information about their location~\cite{drop}. This allows us to infer the location using airport codes, country codes as suffixes, and city names. This combined with the latency measurements allows us to come up with country name}
%For manual analysis, we manually performed traceroutes to the IP addresses and decided the locations based upon next-hop routers, RTT and the hostnames of the routers.\drc{This is still not convincing.} \muzammil{Hostnames generally contain hidden information about their location\cite{drop}. This allows us to infer the location using airport codes, country codes as suffixes, and city names. This combined with the latency measurements allows us to come up with country name}
%
%We also included

%\subsection{Geolocation Sources}
%\label{sec:geolocation_sources}

\subsubsection{Feature Selection}
This section describes the features we select from unreliable data sources for training our classifier. We summarize the selected features and their individual accuracy for geolocating routers in Table~\ref{tab:classifier_features}.

%\begin{table*}[]
\begin{table}[]
\tabletextsize
\setlength\tabcolsep{4pt}
	\centering
	\rowcolors{2}{gray!10}{white}
	\begin{tabular}{ l | l | l | l | l}
		%\hline
		\makecell{\textbf{Information} \\ \textbf{Source}}								& \textbf{Cost}			& \textbf{Precision}				& \makecell{\textbf{Selected} \\\textbf{Features}}				& \textbf{\makecell{Accu-\\racy}}																						\\ \hline
		IP address									& N/A					& N/A								& IP address				& N/A																									\\
		\makecell{CAIDA AS~~ \\Rank}		                 & Free                     & N/A						 		 	& \makecell{ISP size (\# ASes), \\ customer cone \\ (\# of IPs \&  prefixes)}		& N/A             \\
		WhoIS										& Free					& N/A								& \makecell{AS: name,\\  number, country	}																							& 67\%\\
 & 					& 								& \makecell{ISP: name, city, \\  region, country}																							& 69\%\\
		DDec											& Free					& City							& Country	 & 83\%																									\\
		DB-IP										& Free					& City							& Country							& 69\%																									\\
		\makecell{Maxmind \\ GeoLite2}                     & Free                     & Country						 & Country                  				& 71\%																			\\
		APIgurus                     		 & Paid                     & City						 & Country                  					& 70\%																		\\
		IP2Location                      	 & Paid                     & Country						 & Country                  				& 81\%																			\\
		IP Info 	                      	 & Paid*                    & City						 & Country                  						& 71\%																	\\
		\multicolumn{4}{l}{* provided for free for our research}																					\\
		% \hline
	\end{tabular}
	\caption{\textbf{Classifier features} {\sl and their individual accuracy for predicting router country. While some data sources offer high overall accuracy, we show in Fig.~\ref{fig:results_classifier_comparison_accuracy_ensemble} that this accuracy is limited to a small number of countries that are overrepresented in our dataset.}}
	\vspace{\postfigspace}
	\label{tab:classifier_features}
\end{table}
%\end{table*}

%The features used to both train and classify the IP addresses to countries were collected from a set of geolocation sources\drc{What does this mean?} as well as the WhoIS information available publicly.\drc{You list two other sources of data below that are not listed here.}  
%The geolocation services used were IPInfo\cite{ipinfo}, Maxmind Geolite2 Database\cite{maxmind}, DB-IP\cite{dbip}, EurekAPI\cite{eurekapi}, and IP2Location\cite{ip2location}. 
%Other sources of information included the WhoIS information, and AS Rank\cite{asrank} and DDEC\cite{ddec} by CAIDA. 

The feature we use from IP geolocation services and hostname parsers (IPInfo~\cite{ipinfo}, Maxmind Geolite2~\cite{maxmind}, DB-IP~\cite{dbip}, APIgurus~\cite{eurekapi}, IP2Location~\cite{ip2location}, and DDec~\cite{ddec}) is the country corresponding to the geolocation for each IP. Further, we use the registered country/countries for each IP's ASN and ISP (which may own multiple ASNs) based on \emph{whois} data.

% are some of the publicly available geolocation sources.\drc{How many others are there? Why didn't you use some of them?} They convert an IP address to a geographical location. All sources except IPInfo\cite{ipinfo} provide different granularity of accuracy, ranging from a city (or sometimes a zip code) to a country with a varying cost and query limits. IPInfo\cite{ipinfo} provides the same service with varying query limits only.  All these sources boast a very high accuracy for geolocation; however, as discussed in section \muzammil{cite section 5}, these sources have a significantly lower accuracy for router locations.\drc{Be specific, use numbers we measured.} We use the countries predicted by these geolocation sources as the one of the features for the classifier.
%
% is service by CAIDA that utilizes machine learning on hostname parsing (where one is available) and tries to find the location of the hostname based on airport codes, city codes and other properties. Specifically, DDEC\cite{ddec} combines DRoP\cite{drop} and undns\cite{undns} to infer location solely from hostnames\drc{What does DRoP contribute and what does undns contribute to DDEC?}. \drc{You did not say what features is used here.}\muzammil{DDEC\cite{ddec} hostname decoding database that combines hostname parsing rules from \textit{both} DRoP\cite{drop} and undns\cite{undns} to infer a city/country-level location. We used the country predicted by DDEC as feature.}

We include the size of each AS as features, under the hypothesis that larger ISPs are more likely to span multiple countries, while smaller ones are likely to be within a single country. We define size using the number of ASes in each ISP, the number of IP addresses, and the number of routable prefixes contained in an ISP's customer cone (using \cite{asrank}). 

%A full list of all the features used is summarized in Table~\ref{tab:classifier_features}. 
We evaluated the \textit{feature importance} (using Gini importance \cite{louppe:2013:gini-importance}) of the features in determining a router's country, and found that IP2Location~\cite{ip2location} was the primary determinant of the predicted country for 71.0\% decisions. This was followed by the IP address (6.70\%), and country predicted by DDec~\cite{ddec} (6.44\%). The ordered list of features by importance is available in Table~\ref{tab:feature-importance-all-default} in Appendix~\ref{sec:appendix:feature-importance}.

%\muz{There was a decrease in average accuracy by quarter (1\%) of a standard deviation, when any of these features were removed, except IP2Location. Removing it caused a decrease of 20\% in accuracy}

%is the information provided by CAIDA about the size of each ISP (in terms of customer cone size). We use this information so that larger ISPs and ASes can be differentiated from smaller ones since larger ISPs tend to span multiple countries while smaller ISPs might located within the same country.\drc{Not quite right. We do not use this information directly in this way. We hypothesize that the size of an ISP may have an impact on how well it is geolocated, and make this available to a classifier for prediction. Have you shown that this hypothesis is true? If we can not show it, we probably should not include it.}\muzammil{No, we've not proven this hypothesis.} 

%The  provides a summary of the sources and the individual features of each source used to classify the IP addresses to countries.

\subsubsection{Classifier Selection and Training}
\label{subsubsec:methodology-classifier}

When implementing our machine-learning approach to predicting countries where routers are located, we identified two key 
challenges: selecting the type of classifier(s) to train and determining the information provided for training. 

\para{Classifier Selection.}
Our primary goal is to identify a classifier that can provide high accuracy and precision for predicting router country. A secondary 
goal is to use a classifier that is human-interpretable so we can ascertain why it performs well.   \muz{cut last 5 paras?}

\begin{table}[]
\tabletextsize
	\centering
	\rowcolors{2}{gray!10}{white}
	\begin{tabular}{ l | l  }
		%\hline
		\textbf{Classifier} & \textbf{Accuracy(\%)} \\ \hline
		Decision Tree                             & $88.45 \pm 5.11 $               \\
		Random Forest                             & $84.38 \pm 5.4 $               \\
		Extremely Randomized Trees                & $84.41 \pm 3.91 $                \\
%		Decision Table                               & $89.16 \pm 4.54 $              \\
		1-Nearest Neighbor                         & $58.19 \pm 7.79 $                \\
		AdaBoost                                  & $32.05 \pm 9.39 $                \\
		Linear SVM                               & $15.83 \pm 8.67 $              \\
		Naive Bayes                               & $12.07 \pm 5.09 $              \\
		% \hline
	\end{tabular}
	\caption{\textbf{Classifier Accuracy} {\sl using 10-fold cross validation. Decision trees provide high accuracy.}}
	\vspace{\postfigspace}
	\label{tab:result_classifier_accuracy}
\end{table}

%\subsubsection{Accuracy for different classifiers}
%\label{sec:results_classifier_comparison}
We evaluated all available classifiers in \textit{scikit-learn}~\cite{scikitlearn:python} to find ones with high accuracy. These include tree-based classifiers and ensembles (\eg Decision Trees and Random Forest), clustering algorithms (K-Nearest Neighbors), Support Vector Machines (SVMs), Naive Bayes, and adaptive approaches (\eg AdaBoost) as listed in Table~\ref{tab:result_classifier_accuracy}.

We evaluate accuracy on the entire labeled dataset using 10-fold cross validation; \ie we sort the labeled data randomly and evenly divide it into 10 folds (disjoint groups). We test the accuracy of the classifier by training it on nine folds and testing on the one that was not in the training set. This is repeated 10 times, one for each fold to exclude from the training dataset, and the average accuracy (as well as standard deviation) are presented in the second column of Table~\ref{tab:result_classifier_accuracy}. We refer to such classifiers, trained on the entire training dataset, as \emph{default} classifier.

There are clear winners and losers. Decision trees and their ensembles perform well (with similar accuracy), while clustering approaches and AdaBoost do not. However, this finding is potentially biased by the sample size per country in our dataset, and may simply indicate that the classifiers do well only for the countries with a large sample size.

\begin{figure}[t]
    \centering
    \begin{subfigure}[t]{0.48\linewidth}
		\includegraphics[width=\columnwidth]{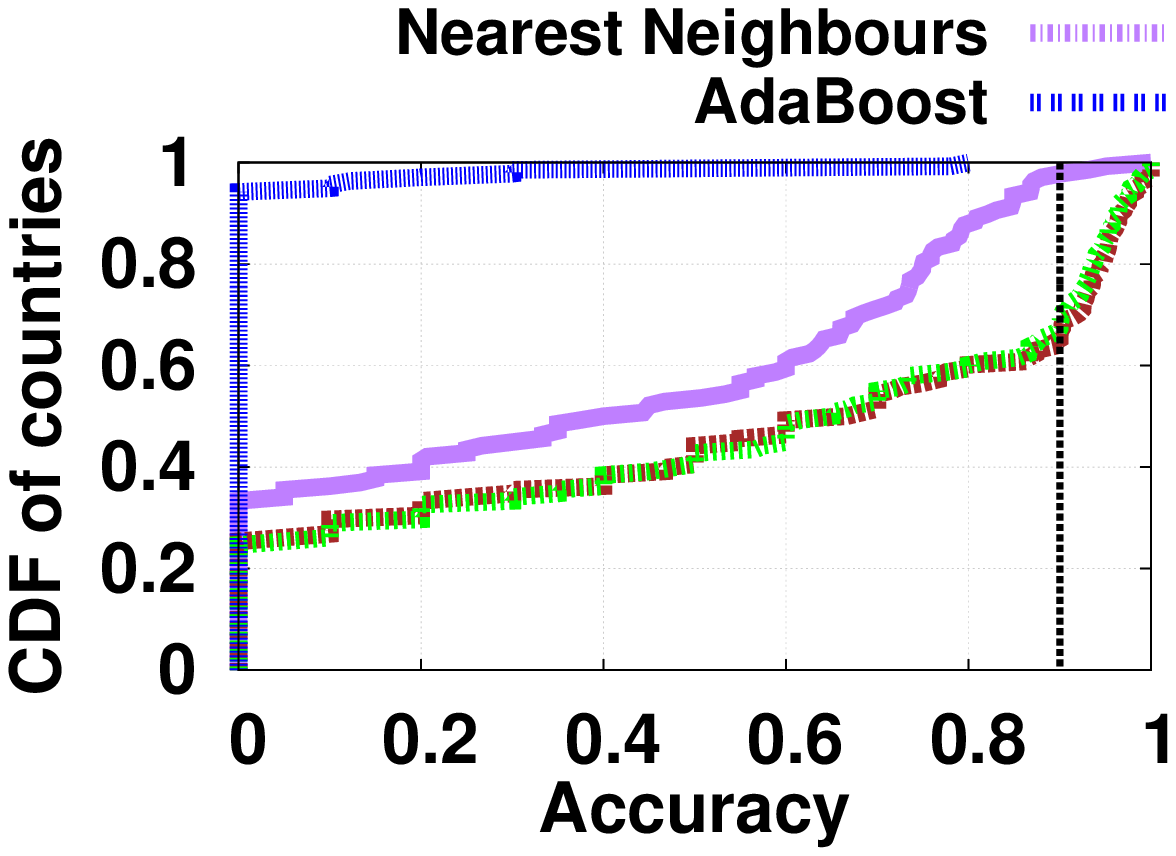}
		\caption{Random}
		\label{fig:results_classifier_comparison_accuracy_all}
    \end{subfigure}
    ~
    \begin{subfigure}[t]{0.48\linewidth}
		\includegraphics[width=\columnwidth]{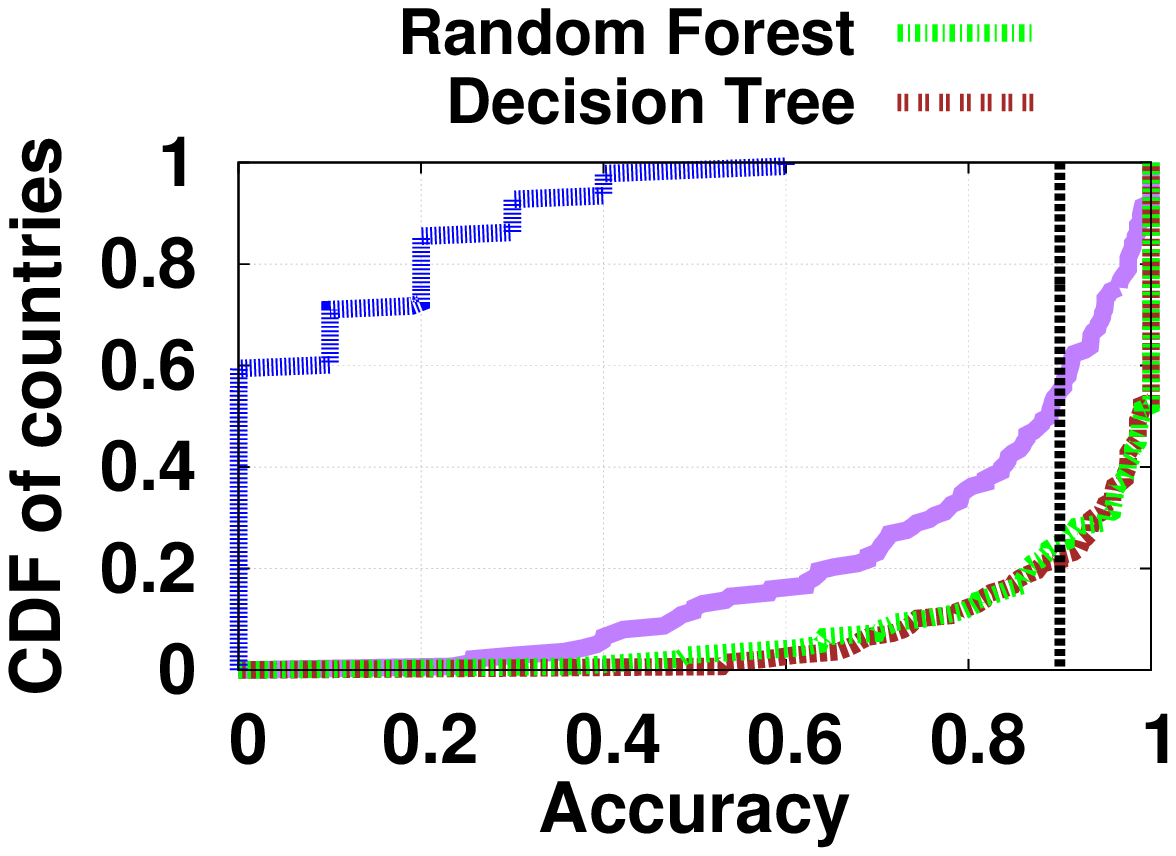}
		\caption{Equal}
		\label{fig:results_classifier_comparison_accuracy}
    \end{subfigure}
    \caption{ \textbf{Classifier Comparison.} {\sl CDF of the fraction of accurately predicted router locations per country, for each classifier type that performed well on the overall dataset.  
		When training the classifiers using (a) randomly sampled number of instances, and (b) equal number of instances in each country, decision trees and forests provide the best per-country accuracy.}}
    \label{fig:results_classifier_comparison_accuracy_bth_gph}
    \vspace{\postfigspace}
\end{figure}

To investigate this, we compare the top-performing classifiers from Table~\ref{tab:result_classifier_accuracy} in terms of the accuracy \emph{per country}. For each classifier type, we train one instance using data from a uniformly  \textit{random sample} of training examples, and a separate instance using an \emph{equal number of training examples per country} (\ie by oversampling from countries with few routers and undersampling from those with large numbers of routers)\footnote{Note that we  evaluate additional training dataset selection approaches to understand classifier sensitivity in Section~\ref{sec:evaluation}.}. Figure \ref{fig:results_classifier_comparison_accuracy_bth_gph} shows our per-country accuracy results. We found that Decision Trees and Random Forests substantially outperform all other classifiers, regardless of sampling method (for details, see Appendix~\ref{sec:appendix:classifier-selection}). Given that decision trees perform well and are easy to interpret,\footnote{As expected, we find that the decision-tree nodes encode information about which database is most accurate for a given IP address.}  we selected decision-tree based classifiers for our implementation. \emph{Note, however, that no individual classifier performs well for \textit{all} countries.} This motivates the need for ensemble approaches as described below.

%In summary, decision trees and random forests provide high overall and per-country accuracy.
%\drc{Did you actually check for this?} \muz{yes, IP2Location.}

%These figures show that a training data bias affects the per-country accuracy for different countries and 
%single one-size-fits-all classifier doesn't capture the broad scope of instances that are possible.\drc{I can't quite follow the second point, nor do I think you have shown it here.}
%\muz{added ends}

\para{Training an \emph{ensemble} of classifiers.}
We have established that the training dataset used for classifiers can have a significant impact on accuracy, particularly when there is bias in the dataset. 
Thus, it is important to take these biases into account in \sysname to provide high accuracy across a wide range of countries that Internet paths may visit. 

A key challenge is that we cannot know \emph{a priori} how biased our training dataset is relative to an arbitrary set of traceroutes needing router-country predictions.  Thus, instead of attempting to provide a ``one-size-fits-all'' classifier trained with some subset of training data, we chose to implement an \emph{ensemble} of classifiers, each of which is trained using complementary approaches to sampling our dataset. In this model, each classifier in the ensemble predicts router countries independently and the ensemble returns a set that is the union of these countries. As we discuss in the following paragraphs, we use RTT latency with SoL constraints from traceroute data to eliminate infeasible countries from this set. If the set still contains more than one country, \sysname can optionally return the set or report that it was unable to isolate a single country for the router. In our evaluation, we show that \sysname returns exactly one country 88.1\% of the time, and zero countries only 3.1\% of the time. %\muz{added} \sysname itself predicts only a  \emph{single} country or fails to narrow down the list where SoL constraints are not available or fail to reduce to set of countries.

%Our ensemble consists of 3 classifiers. One classifier simply uses the entire labeled dataset. For the remaining classifiers, we train instances of each type, each using a different subset of data to avoid bias in the dataset selection process. One of the classifiers uses a fixed number of training instances per country, under the assumption that doing so will reduce the impact of country-level bias in the dataset.

%\drc{I don't understand how there are two per \# of samples} \muz{I didn't do an analysis on the optimal number for variants for each classifier. Most of these number of instance per countries were usually a little over half of their number in training data that's why I chose two.} 

Our \ensemble consists of multiple classifiers.\footnote{We tested 23 classifiers, not shown due to space limitations, and used the 4 that provide highest accuracy gain in our \ensemble.} % (Table~\ref{tab:classifiers_ensemble_included}). 
One classifier simply uses the entire labeled dataset. For the remaining classifiers, we train each instance using a different subset of data to mitigate bias in the dataset. One of the classifiers uses a fixed number of training instances per country, as we found that doing so will reduce the impact of country-level bias in the dataset. The remaining two classifiers are trained using empirically derived sets of training instances. For this, we investigate the marginal impact on accuracy from adding individual training instances for each country.  One approach uses this to find the minimal set of training instances that maximizes the accuracy for each country (which we refer to as \emph{maximum accuracy}). The other approach finds the point at which adding more instances has diminishing returns on accuracy (\ie the second derivative of the accuracy vs. number of instances curve is zero), which we refer to as \emph{knee} (of the curve).

\para{SoL constraints and iterative refinement.}
The \ensemble classifier predicts one or more countries for each router. To improve the accuracy and precision of the ensemble, we use traceroute data in the following way. 

First, we use RTT latencies associated with each traceroute path that includes the targeted router, $r$. For each source $s$ in the set of sources $S$ with a path that includes targeted router $r$, we find the minimum RTT delay between $s$ and $r$. We conservatively find the distance $d$ between $s$ and $r$ using the empirically derived propagation delay from Laki \etal~($0.47c$)~\cite{Laki:2009:path-latency}, then identify the set of permissible regions that $r$ can be located in using the geodesic circle $c_s$ centered at $s$ (geolocated using a geolocation service) with radius $d$. When there are paths from multiple sources containing $r$, we identify the set of permissible regions for $r$ as bounded by the intersection of circles $c_s$ for all $s$ in $S$. 

After identifying the permissible region where $r$ may be located, we identify the countries intersected by region. Finally, we adjust the set of predicted countries by finding their intersection with the permissible countries. At the end of this step, a router is assigned to zero\footnote{The intersection of permissible regions $c_s$ may be an empty region, or permissible regions may not include any predicted countries.} or more countries.

We incorporate two optimizations to improve coverage and efficiency. First, we use lists of router aliases~\cite{keys-midar} to ensure that we can intersect paths visiting different IP addresses on the same router.  Second, we ignore large latency measurements (which lead to intersection regions larger than a country).
%and inconsistent vantage points\drc{inconsistent how? And how do you determine which inconsistencies should be thrown out?}\muz{it was inconsistent in terms of the VP location. The location information was not available for one Ripe Atlas probe}. 
Namely, we use only measurements with an RTT less than $100$ milliseconds to ensure that the diameter of $c_s$ is less than $40$\% of the circumference of the earth in any direction.

% DRC: Removed details here, we could always actually do geodesic intersection
%We use bounding boxes surrounding (geodesic) circles to reduce the computational cost of finding intersecting regions. We assume that queuing delays are not substantial in our traceroute dataset. 
%In cases where this is not true, each $c_s$ affected by queueing will be too conservatively large, leading to low precision (too many countries may be included) but not a loss of accuracy (a permissible country will not be eliminated from $c_s$).

For cases where the latency from $r$ to all $s$ in $S$ is large (\ie our vantage points are far from the targeted router), we use an iterative refinement approach as follows. Given a path measured from $s$ containing $r$, we find all \emph{landmark routers} $l$ with IPs that can be located with high precision (\eg reliable geolocation sources from \S\ref{sec:dataset}). 
%This includes IPs contained in our set of reliable geolocation data sources described in Section~\ref{sec:dataset} and those that have been located with high precision using the above approach. 
We then estimate the latency between $l$ and $r$ as $\frac{1}{2}*|rtt_{s-r} - rtt_{s-l}|$, where $rtt_{s-r}$ is the RTT between $s$ and $r$ and $rtt_{s-l}$ is the RTT between $s$ and $l$. 
In other words, we assume that the path from $l$ back to $s$ is the same as $r$ back to $s$, and find the average additional delay from $l$ to $r$ and $r$ to $l$ (which is equal to or larger than the smallest unidirectional delay on the round-trip path). Because the above path assumption may be violated in practice, we use this approach only when no other information is available to reliably isolate a router country. %Validating these assumptions using a tool like Reverse Traceroute~\cite{revtr} and one-way delay measurements is a topic for future work.

% additional delay from $s$ to $r$ and $l$ is \emph{symmetric} along the forward and reverse path, and

%\muz{added}
%Our assumption of using a \emph{symmetric} delay doesn't violate the \emph{asymmetry} of the internet paths since the \emph{symmetric} delay assumes the \emph{average} of the forward and reverse path latency. This \emph{average} of the latency values is always an overestimate of the actual distance that could have been traveled by the packets as it is an \emph{average}, and the average is always greater than the smaller of the two one-way latency values.
%\muz{end}

After this step, each router $r$ is mapped to a permissible set of countries (or the empty set). We use this as a new label to retrain the \ensemble classifier to improve its predictions. We repeat the above process until the set of countries mapped to routers stabilizes.

%aliases
%assumptions about symmetric delays
%assumptions about queuing
%bounding boxes instead of discs
%
%
%The SoL system is similar to Alidade\cite{alidade}.\drc{Be more specific about how it is the same and how it is different.} We select a few landmark locations\drc{what is a landmark?} based on certain attributes of the IP address\drc{which attributes? You should say, "the following attributes" to avoid this question. Also, why these attributes?}. These attributes are the IXP location, location from DDEC \cite{ddec}, location based on hostname parsing rules (written by us\drc{needs detail}), and location inferred from airport and city codes embedded in the hostname\drc{how is this different from the previous two?}. If all these sources provide a similar\drc{same country?} location, that location is declared a landmark. Using these landmarks we use 2/3 x speed of light\drc{why 2/3?} with RTT\drc{phrasig is strange} to find the maximal possible distance of each possible IP address. Using the intersection of these regions we come up with a country (or a set of countries) where this IP address might be located.  These intersections are saved and used again to location next hops, iteratively.
%
%To optimize speed of intersection in our approach, we use rectangular intersection, instead of circular intersection. 

\subsection{Online Prediction}

%\muz{modified}
%The online prediction system is developed as a web service in Flask, a web framework in Python. 
A user submits a traceroute (or an IP address)  to the \sysname web server using an online submission form or API request. \sysname parses the traceroute to extract IP addresses and RTTs, then uses its classifier to predict the countries where these IPs are located, and applies SoL constraints to eliminate any infeasible countries. The output is  the filtered list of IP-to-country mappings. The online system predicts each IP address to \emph{up to two} countries or returns an error for that IP address. The detailed predictions of countries from the \ensemble classifier, the list of countries filtered using SoL constraints, and a complete list of countries predicted by \sysname can be accessed using the public API. 

The online prediction system, user guide, and the API are publicly available at \url{https://passport.ccs.neu.edu}

\section{Evaluation}
\label{sec:evaluation}

This section addresses key questions regarding the effectiveness and efficiency of our approach. First, we conduct microbenchmarks on our classifier design and implementation. Next, we demonstrate the advantage of incorporating SoL constraints and iterative refinement. We then show that our approach incurs reasonably low overhead to train and is suitable for online prediction. Last, we compare our approach with alternatives. 
%Last, we look at the analysis of paths by our approach over a period of time.

\subsection{Methodology}

We evaluate our classifier design and implementation using the same dataset and methodology presented in Section \ref{subsec:offline-training}. This contains a set of router IP addresses for which we have ground truth labels. To evaluate our system against IPs not in our training dataset, we conduct additional traceroutes (using a similar methodology) as specified in Section~\ref{subsec:constraint-refinement}.

%This ground truth information spanned over 130 countries in the world and the some of the manual information was chosen from IP addresses seen belonging to most important ISPs in majority of countries.
%This selection of ground truth information allows us to diversify our training dataset and be able to maximize the number of predicted countries, since machine learning has a limitation of only being able to predict information that is previously seen.

\subsection{Classifier Analysis}
\label{sec:results_classifier}

In this section, we answer the following key questions:

\begin{packeditemize}
\item \emph{How well do simple approaches work at predicting IP location?} None is particularly accurate, especially when considering \emph{per-country} accuracy.
\item \emph{Are all the unreliable geolocation sources similarly bad, or are there cases where at least one correctly predicts locations most of the time?} Usually at least one unreliable source can correctly predict location, opening the door to classifier-based prediction.
%\item \emph{Can we train an ML classifier to use this observation to yield higher accuracy than any individual source?} Yes.
\item \emph{Is any one classifier sufficient to optimize accuracy?} No, an \ensemble provides the best results.
\item \emph{Is the machine-learning approach both accurate and precise?} Yes, in the vast majority of cases our approach identifies one country for a router.
\end{packeditemize}

%We now present the analysis that allowed us to answer these questions. 
The following subsections describe which ML classifiers work best individually and how we combine them to provide higher accuracy. All these results present average statistics using 10-fold cross validation. 

\subsubsection{Overall accuracy}
\label{subsec:overall-accuracy}
We start with the accuracy of our final \ensemble classifier as described in Section \ref{subsubsec:methodology-classifier}, using the labeled data from Section \ref{sec:dataset}. Accuracy is defined in terms of the fraction of router-to-county predictions that are correct according to ground truth labels. We exclude cases where multiple countries are predicted. 

%If a set of countries is predicted (as is the case with our \ensemble) we consider it to be correct if one of the predicted countries matches the ground truth country for a router. 

\begin{figure*}[!htb]
\minipage[t]{0.32\textwidth}
	\includegraphics[width=1.0\columnwidth]{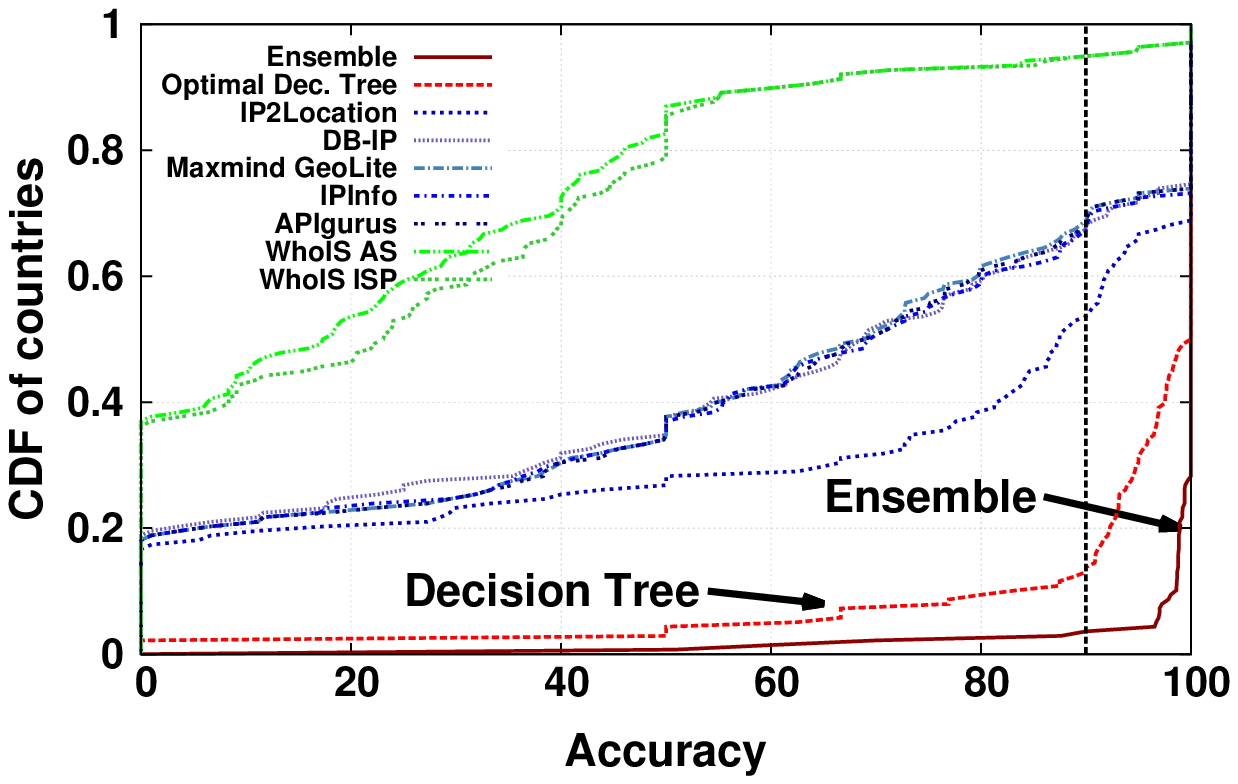}
	\caption{\textbf{Comparison of our \emph{ensemble} approach with alternatives}, using a CDF of per-country router location-prediction accuracy. 
		{\sl The \ensemble curve (bottom-right) has a high accuracy for most countries, while traditional approaches have high accuracy only for a few countries.}}
	\vspace{\postfigspace}
	\label{fig:results_classifier_comparison_accuracy_ensemble}
\endminipage\hfill
\minipage[t]{0.32\textwidth}
	\hbox{\hspace{-0.5em}\includegraphics[width=1.07\columnwidth]{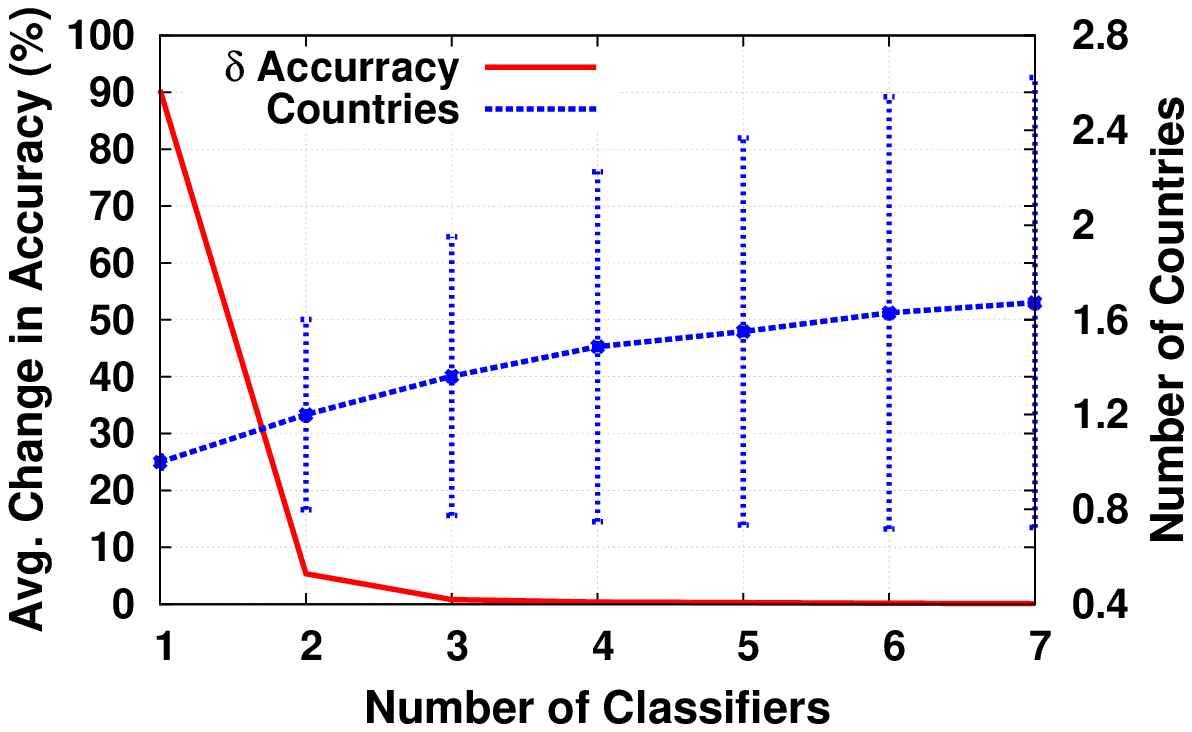}} 
	\caption{\textbf{Change in average accuracy for all countries for each new classifier}. 
		{\sl We find that four classifiers are sufficient for high accuracy. By limiting the classifiers in the ensemble, the predictions are more precise (\ie include fewer countries). }}
	\vspace{\postfigspace}
	\label{fig:runme_optimal_classifiers_delta}
\endminipage\hfill
\minipage[t]{0.32\textwidth}%
	\includegraphics[width=1.0\columnwidth]{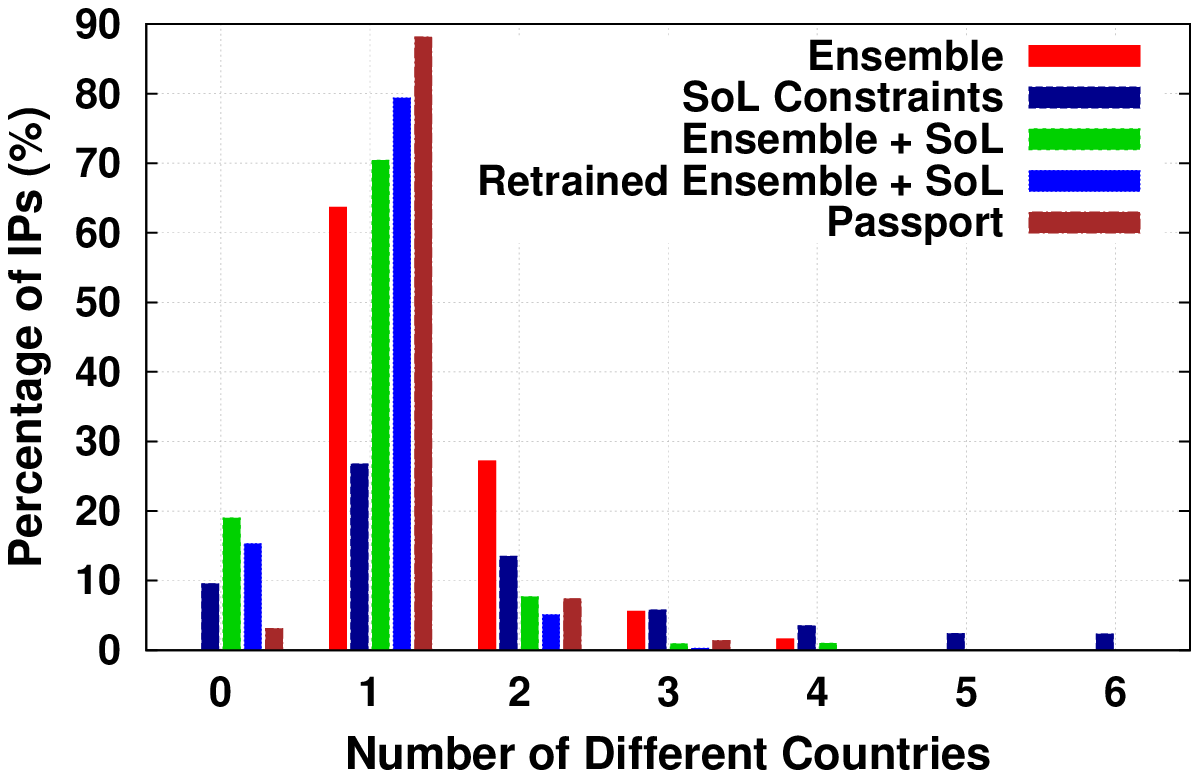}
	\caption{\textbf{Impact of using speed of light (SoL) constaints}. 
		{\sl By incorporating SoL constraints to rule out infeasible router locations, then using this information to retrain our classifiers, our system predicts exactly one country \pctPredSingleCountry of the time, compared to the initial classifier that does so only 63.7\% of the time. }}
	\vspace{\postfigspace}
	\label{fig:results_iterative_analysis_stats_country}
\endminipage
\end{figure*}

We compare the performance of our \ensemble with different \geolocation services in terms of the fraction of routers in each country that are correctly predicted (singleton set). This is plotted in Figure~\ref{fig:results_classifier_comparison_accuracy_ensemble} using a CDF where the y-axis represents countries and x-axis is the fraction of a country's routers predicted correctly. Curves closer to the bottom and right edges indicate higher accuracy. 

Our \ensemble substantially outperforms all other approaches---it achieves 90\% or better accuracy for 96.5\% of countries in our dataset. By comparison, individual decision trees can achieve the same level of per-country accuracy only for 61\% of countries. Worse, the most accurate IP geolocation service, IP2Location achieves this only for 46.5\% of countries and \emph{whois} registry information for only 5\% of countries. 

In terms of implications, it is clear that machine-learning approaches are able to synthesize individually inaccurate geolocation sources to more reliably predict router geolocations. Further, the accuracy for IP geolocation services and \emph{whois} data do not extend beyond a small fraction of countries, \emph{meaning any conclusions about geopolitical properties of Internet paths using such data (\eg \cite{edmundson:ran,shah:bgpdetours}) are highly likely to be incorrect.}

%We use a CDF of the countries and accuracy of all the services for the respective countries in the Figure \ref{fig:results_classifier_comparison_accuracy_ensemble}.\drc{Does this include SoL information?} The curves closer to the bottom right have better accuracy.  The x-axis represents the accuracy while the y-axis is the CDF of the countries (in out training dataset). The graphs show that 96\% of countries have an accuracy of over 90\% using the \ensemble while using a traditional 63 \% countries have attain a 90\% accuracy using a simple classifier, 46\% can be correctly predicted for similar accuracy by using IP2Location, but only 4\% can be predicted correctly with atleast a 90\% accuracy using the WhoIS information. \drc{Also mention median accuracies}
%
%It appears that the accuracy of the geolocation services is relatively lower as compared to the classifiers in general. The accuracy for IP2location\cite{ip2location} is comparatively higher than other geolocation services and WhoIS information has the lowest accuracy.

\subsubsection{Microbenchmarks and Classifier Sensitivity}
\label{sec:results_classifier_training_sensitivity}

We now investigate the individual components of our \ensemble classifier and their sensitivity to the training dataset. 
For each country, we select the number of training instances that either (1) provide maximum per-country accuracy (\emph{maximum accuracy}), or (2) at the point where increasing instances provide diminishing returns in accuracy (\emph{knee}), as discussed in \S\ref{subsubsec:methodology-classifier}. 

The outcome depends on how we allocate training instances to \emph{other countries}, so we investigate three schemes using three schemes: using the same number of samples for other countries as in the target country (\emph{balanced}), double the instances in other countries as in the target country(\emph{unbalanced}), and a random number of samples for other countries (\emph{random}).

Overall, we find that \emph{random} performs systematically worse, while \emph{maximum accuracy} and \emph{knee} approaches on both \emph{balanced} and \emph{unbalanced} sampling methods yield similar results. The details on how we select training instances to achieve high accuracy are further described in Appendix~\ref{sec:appendix:results_classifier_training_sensitivity}.

\subsubsection{Ensemble Construction}
\label{subsubsec:ensemble-construction}
% \drc{Start the paragraph by stating (1) what you have learned by the end of Fig. 7, (2) what questions remain unanswered, and (3) how you answer them.} 
In Section~\ref{subsubsec:methodology-classifier}, we described several techniques for training classifiers in addition to the \emph{default} classifier. We now investigate how to combine these classifiers to optimize for accuracy and precision.

Figure \ref{fig:runme_optimal_classifiers_delta} shows the change in accuracy for all size-\emph{n} sets of classifiers with the number of classifiers in the set (x-axis) against the average increase in the accuracy (y-axis) and the number of countries predicted by the set of classifiers (y2-axis). 
The figure shows a diminishing return on accuracy improvement as more classifiers are added. At $x = 3$, the increase in average accuracy is roughly $1.1\%$ while it decreases to $0.3\%$
 for $x=4$ and beyond this point the increase  in accuracy is negligible.

We pick the first \numClassifiers classifiers to form our \ensemble classifier. 
It contains \emph{default}, \emph{balanced} dataset bias with \emph{knee} sampling approach, \emph{unbalanced} with \emph{maximum accuracy} approach, and one classifier with equal number of training instances for all countries. 

\subsubsection{Precision}
\label{sec:results_classifier_precision}

A key question is whether the high accuracy in the \ensemble classifier is due to one correct country being predicted for a router, or rather that the correct country is one of a large set of countries returned.
%To evaluate this question of precision, we computed the distinct number of countries and continents returned by the classifier for each router in our dataset.
To evaluate the precision, we computed the distinct number of countries and continents returned by the classifier for each router in our dataset.
The \ensemble predicts a single country for 65.7\% of router IPs, while it predicts a single continent for 82.4\% of router IPs (see Appendix~\ref{sec:appendix:results_classifier_precision}). Thus, the \ensemble alone is accurate but not precise; further, due to the spread in continents predicted, this precision is not sufficient to reason about geopolitical implications of Internet paths.
%In the next section we demonstrate how we use SoL constraints to improve precision such that the vast majority of the time (\pctPredTwoCountry of IPs) a router is predicted to be located in one or two countries.
In the next section, we demonstrate how we use SoL constraints to improve precision such that the vast majority of the time (\pctPredSingleCountry of IPs) a router is predicted to be located in one country.

%
%, using a CDF in Figure~\ref{fig:results_classifier_precision}. A point in each curve represents the fraction of IP addresses (y-axis) for which at least $x$ countries were predicted (x-axis). 

%The graph shows 

%\begin{figure}[tb]
%	\centering
%	\includegraphics[width=1.1\columnwidth]{results_classifier_precision}
%	\caption{\textbf{Precision}. 
%		{\sl The total number of countries mapped by the \ensemble classifier for each unique IP address.}}
%	\vspace{\postfigspace}
%	\label{fig:results_classifier_precision}
%\end{figure}

\begin{comment}

* Assuming this talks about the precision of the ensemble.

* Graph

- A graph that is similar to CDF

- x= country id (sorted by the number of countries predicted)

- y = the number of different countries predicted by the ensemble classifier for that given country.

- graphs: only one.(maybe we can add more by choosing say, top 3 sources of accuracy in ensemble (from previous subsubsection) vs all, from the table above) 

- reason: - 

\drc{No. A CDF over routers, x-axis is \# of countries being mapped to.  x axis- number of countries. y axis- cdf of routers. each point is the number countries a router is mapped to. it'll be a step function.}

\end{comment}

\subsection{Constraint-based Refinement}
\label{subsec:constraint-refinement}

In this section, we discuss how we use RTT latencies measured from traceroute paths to substantially improve the precision of the \ensemble classifier. 
Specifically, we use SoL constraints as described in Section~\ref{subsubsec:methodology-classifier} to rule out infeasible countries from the set predicted by the \ensemble classifier. Further, we leverage these observations to train a classifier on the subset of feasible countries predicted by the \ensemble to improve precision for routers without sufficiently narrow SoL constraints imposed by RTT latency.    

\para{Dataset.}
%\drc{Why are we switching datasets? What is the motivation for this? I don't like what I wrote below, it's weak.}
To analyze the impact of SoL constraints, we need a large set of traceroute data from which we can impose substantial numbers of constraints. We thus conducted an additional measurement campaign as follows.  
%for paths between \drc{How many? Which ones?} \muz{87 nodes in 20 countries.} PlanetLab sources and the Alexa top-100 websites as destinations.\drc{why?} \muz{Top-100 global and local websites for 30 countries. Reason: to cover a diverse set of countries and see where our system can perform well, even if no \textit{label} is available for the classifier to predict and then retrain the classifier in order to increase the country dataset (although no new country was found in our analysis)} We use this dataset
We performed forward traceroutes and reverse traceroutes~\cite{revtr} in April, 2017 from 172 PlanetLab nodes in 32 countries to up to four government websites for 190 countries\footnote{We used fewer than four websites if we could not locate four. The limit of four was set due to API limitations by unreliable geolocation sources} (generating 30,248 and 10,477 \emph{successful} traceroutes for forward and reverse traceroutes, respectively). 
To add diversity in terms of sources and destinations, we also used 442,862 traceroutes (311,684 UDP and 131,178 ICMP) from RIPE Atlas from  9,553 probes in 176 countries during the same period. 
These measurements yielded at least two intersecting regions for SoL analysis for 63\% of routers. 
%In addition, 17\% of all routers had high latency (>100 ms) from all vantage points while less than 1\% of hosts received ping responses from multiple VPs but their SoL regions did not intersect. 
%Finally, fewer than 0.01\% routers did not respond to ping measurements.

%\muz{update}17,300 distinct IP addresses (after removing private addresses and source and destination hosts).
%These measurements yielded at least two intersecting boxes for SoL analysis for 28\% of routers. In addition, 13\% of all \drc{of what? all routers or the 35\%?} were in the ground truth location dataset, and 19\%\drc{same question} provided latency from a single source\drc{what does this mean?}. Finally, \muz{update}37\% of hosts received ping responses from multiple VPs but their SoL regions did not intersect. This is likely due to false assumptions about path-latency symmetry along forward and reverse paths.\drc{Is it? Why would this explain it, and why is it likely?}

%SoL intersections (either ground truth locations or intersecting SoL boxes).\muz{I meant that 9\% were marked as landmarks, while only 35\% router had atleast 2 or more intersecting boxes from either landmarks or the source.}\drc{This is not the only case of ``sufficient information.'' What about cases where the 1 box intersected exactly one country in the predicted set?}. 
%\drc{Did you use landmarks at all, or just src->router RTT?} \muz{both of them}

The following paragraphs compare the \emph{precision} of several schemes for using SoL constraints.\footnote{Due to lack of ground truth in this dataset, we cannot evaluate accuracy.} First, we compare the precision of our \ensemble classifier to that of SoL constraints alone, then evaluate the combination of the two techniques. Since SoL constraints can offer additional ground truth labels (\eg by ruling out predicted countries or reinforcing correct predictions), we further evaluate the impact of retraining our classifiers with the combined information. We plot a histogram summarizing our precision results in Figure~\ref{fig:results_iterative_analysis_stats_country}, with each bar indicating the fraction of router IPs (y-axis) that are predicted to be located in $x$ countries.

\begin{comment}
\begin{figure}[tb]
	\centering
	\includegraphics[width=1.1\columnwidth]{results_iterative_analysis_stats_country_overall}
	\caption{\textbf{Impact of using speed of light (SoL) constaints}. 
		{\sl By incorporating SoL constraints to rule out infeasible router locations, then using this information to retrain our classifiers, our system predicts exactly one country \pctPredSingleCountry of the time, compared to the initial classifier that does so only 63.7\% of the time.  }}
	\vspace{\postfigspace}
	\label{fig:results_iterative_analysis_stats_country}
\end{figure}

\begin{table}[t]
	\centering
	%\rowcolors{2}{gray!10}{white}
	\begin{tabular}{ l | r | r}
		%\hline
		\textbf{Source} & \textbf{IPs} & \textbf{Countries} \\ 			\hline
		\emph{Union}                 & \emph{22,213}    & \emph{154}            \\
		\ \ \ Ground truth (Table~\ref{tab:ground_truth})                           & \numTrainSize     & \numTrainCountrySize           \\ 
		\ \ \ SoL                                  & 11,308 & 110       \\
		% \hline
	\end{tabular}
	\caption{Ground truth IP geolocations with SoL.}
	\vspace{\postfigspace}
	\label{tab:ground_truth_with_sol}
\end{table}
\end{comment}

\para{Precision of SoL constraints.} 
The leftmost two bars indicate the precision for the \ensemble classifier and SoL constraints. We find that SoL constraints predict substantially fewer countries than the \ensemble, but neither predicts a single country more than 64\% of the time. The combination of these two approaches (``Ensemble + SoL'' bar), where we use SoL constraints to identify which of the predicted countries is feasible, substantially improves precision with almost all of the routers predicted to be located in 1--2 countries.

\begin{figure}[t]
    \centering
    \begin{subfigure}[t]{0.48\linewidth}
        \includegraphics[width=\columnwidth]{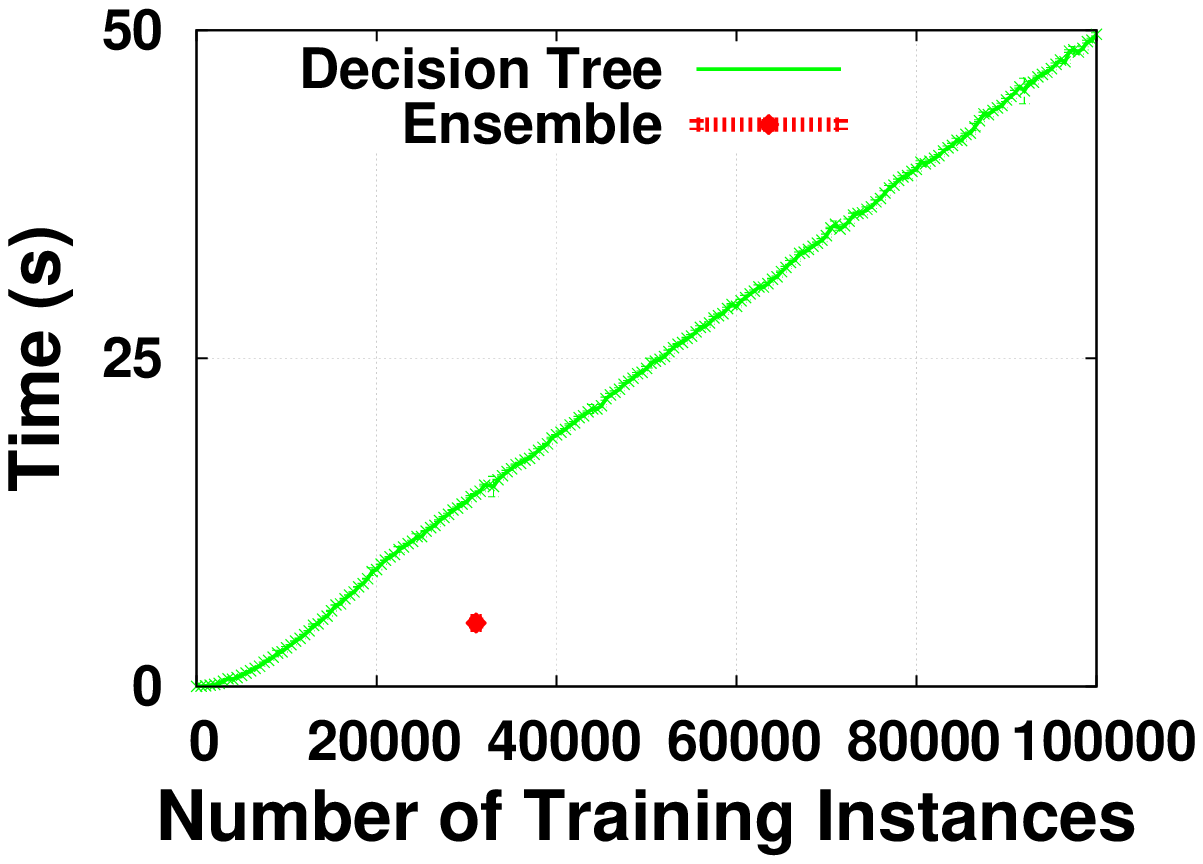}
\caption{Training time}
	\label{fig:results_classifier_time_train}
    \end{subfigure}
    ~
    \begin{subfigure}[t]{0.48\linewidth}
        \includegraphics[width=\columnwidth]{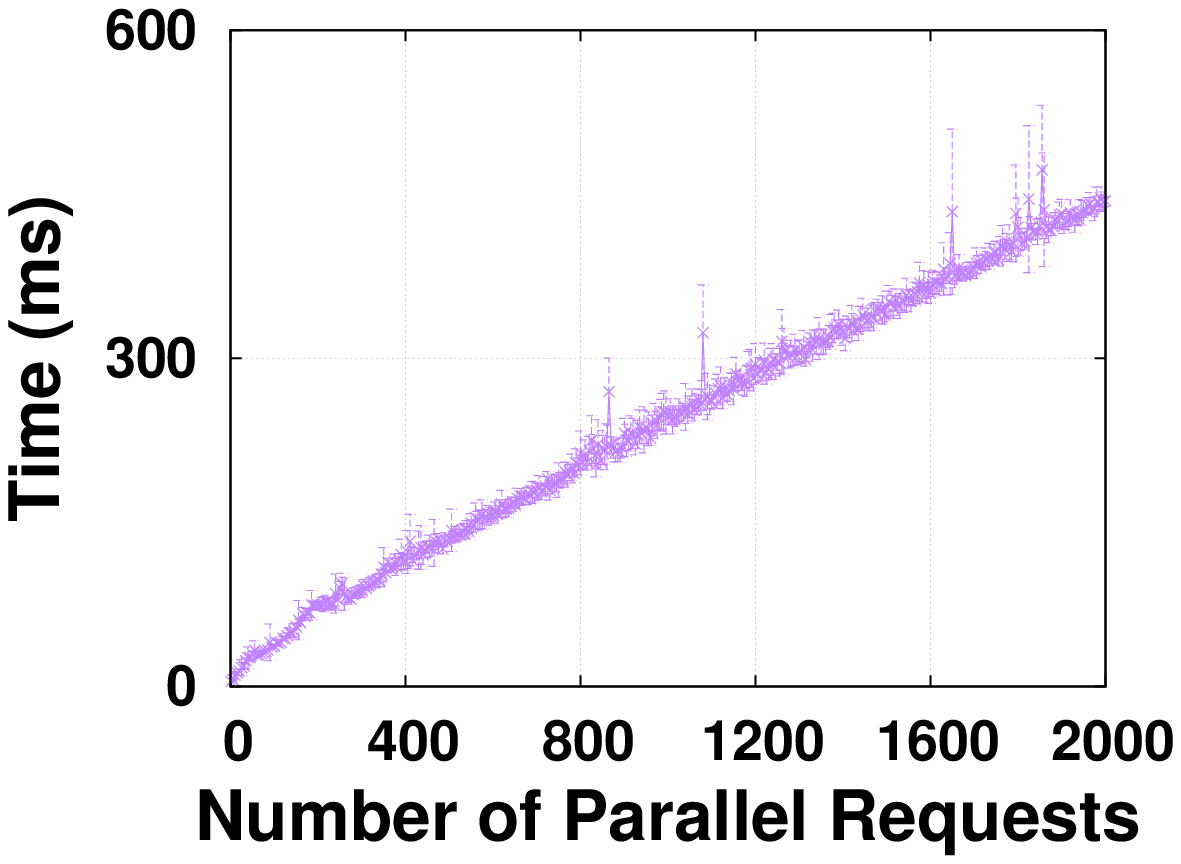}
	\caption{Predicting locations}
		\label{fig:results_classifier_time_test}
    \end{subfigure}
    \caption{ \textbf{System performance.} {\sl Training time is reasonably low and increases linearly with the number of training instances; the online prediction component returns more than 1200 predictions in less than 300\,ms. }}
    \label{fig:performance}
    \vspace{\postfigspace}
\end{figure}

\para{Iterative refinement using classifier retraining.}
%\label{sec:results_feedback}
%In the Figure \ref{fig:results_iterative_analysis_stats_country}, it's still possible to learn the information from the SoL constraints and use this information to retrain the ensemble using a feedback mechanism. 
To further improve precision, we leverage two cases from the previous analysis to retrain our classifier: instances where the SoL constraints identified exactly one country for router (11,308 addresses spanning 110 countries), and instances where the \ensemble classifier was incorrect. We retrain the \ensemble classifier, and using the combination of SoL constraints and the predicted countries by this retrained \ensemble, plotted ``Retrained Ensemble + SoL''. This step substantially improves precision, with 79.4\% of routers being located to a single country and only 5.3\% of predictions containing more than one country. 

The last optimization for precision incorporates cases where the \ensemble predicted multiple countries while the SoL constraints predicted only one country and there was no intersection between the two (showing that the \ensemble was incorrect in predicting countries). In this case, we can ignore the classifiers and \textit{only} use the SoL constraints to predict the country. This is depicted as the rightmost ``\sysname'' bar in Figure~\ref{fig:results_iterative_analysis_stats_country}. 
Putting it all together, our final classifier is able to predict exactly one country for a router \pctPredSingleCountry of the time.

\subsection{Efficiency}
\label{sec:results_classifier_efficiency}

We now evaluate \sysname in terms of time to train the classifiers in the offline component and time to predict in the online component.\footnote{We do not include the variable times necessary to conduct active measurements, though they are on the order of seconds.} All experiments used a  4-core $i7$ processor (4.2 GHz) with $16$\,GB of RAM. 

Figure \ref{fig:results_classifier_time_train} plots the time required to train our classifier (y-axis) as a function of the number of training instances (x-axis), both for the \ensemble classifier and for an individual decision tree.  The graph shows that an individual classifier can be trained with 100,000 instances in approximately 50 seconds. When looking at training the \ensemble of \numClassifiers classifiers used in our system, we find that training time takes approximately 4.8 seconds when trained in parallel (for a combined set of 31,603 instances). Thus, re-training classifiers in the offline phase is not a bottleneck in our system, and if necessary, retraining can be done on the scale of 10s of seconds.

For the online prediction component, we expect users to submit traceroutes and obtain the corresponding countries on the path interactively (\ie within a small number of seconds or less). Figure~\ref{fig:results_classifier_time_test} plots the response time as a function of the number of parallel requests for predictions where measurements are available. The plot shows that under low load the system can make predictions within 10s of milliseconds, serve 350 parallel requests in less than 100\,ms and can serve up to 2,000 parallel requests in less than half a second. Thus, \sysname is sufficiently fast to provide interactive predictions for traceroutes. 

\subsection{Comparison with Alternatives}
\label{subsubsec:results-comparision-alternate}
%In this section, we compare our approach with geolocations provided by other 
%\muzammil{modified}
We now compare our approach with geolocations provided by other 
commonly used services. Unlike the analysis in \S\ref{sec:results_classifier}, 
we use the dataset described in 
\S\ref{subsec:constraint-refinement}, which is larger but does not have 
ground truth labels for router locations. 

\para{Geolocation Databases}
While comparing public databases, instead of determining whether 
%each geolocation service is \emph{accurate}, we evaluate other approaches 
each service is \emph{accurate}, we evaluate other approaches 
in terms of whether they are consistent with \sysname results \emph{for the 
cases where exactly one country is predicted}. While we cannot guarantee 
that all these cases are accurate, we have high confidence in them due to SoL 
constraints and classifier predictions.

Table~\ref{tab:comparison-final} presents the results for several 
geolocation services.\footnote{We attempted to compare with Alidade~\cite{alidade}, but 
the service was not running at the time of writing. 
% from April 2017 and \alidadeSourcesMonthYear.
The authors instead provided us 
with locations provided by Alidade's geolocation services, which are included in the table.} 
We find that all of the geolocation sources yield locations that are inconsistent 
with our data (including SoL constraints), ranging from 7.3--21.5\% of routers in our dataset for April 2017. 
We found that this number increases to 9.1--23.1\% for the same geolocation services with data from \alidadeSourcesMonthYear (not shown), indicating 
that using slightly older geolocation databases (as is the case for Maxmind GeoIP2~\cite{maxmind:geoip2} and IPligence~\cite{ipligence}) yields similar results. 

We evaluate the impact of these inconsistencies when it comes to mapping traceroutes 
to country-level paths and find that a substantial fraction (37.4--66.7\%) of paths are affected.
We find that 0.1--14\% of the inconsistencies occur due to SoL violations, 
and thus are not only inconsistent but \emph{incorrect}.

We found several patterns behind the observed SoL violations. IPInfo~\cite{ipinfo} returns EU as a country for 42\% of their SoL violations (instead of a single country), while Maxmind~\cite{maxmind} incorrectly returns either Switzerland or Sweden for 48\% of the violations. Similarly, IP2Location~\cite{ip2location}, and EdgeScape~\cite{edgescape} return United States as the most commonly mislabeled country, corresponding to 22\% and 11\% of incorrect cases, respectively. IPligence fails to predict a country for over 27\% of violations.

For cases where \sysname and other databases were inconsistent, we sampled and analyzed a subset of cases for manual analysis. 
In the vast majority of cases, we find that \sysname is correct. We sampled and labeled 596 IP addresses from 519 ASes in 146 countries in a manner similar to \S\ref{sec:dataset}. Our results showed that \sysname had the highest accuracy with 96.47\% IP addresses being labeled correctly, followed by IP2Location at 92.3\% and Edgescape at 90.9\%. \sysname outperformed other sources in all continents except Europe, where Edgescape had the highest accuracy  (95.35\%, vs. 93.4\% for \sysname). Appendix~\ref{sec:appendix:geolocation-sources-accuracy-comparison} provides an in-depth comparison of the geolocation sources with \sysname.

%\muzammil{added}
\para{Comparison with active measurement systems.}
A key advantage of \sysname is that it can accurately geolocate router countries with few (or no) additional measurements beyond a traceroute. 
To demonstrate this, we investigated the number of vantage points required to issue ping measurements to determine whether a predicted country is feasible according to SoL constraints. We explored a greedy approach (use VPs with the lowest RTT to a router) and a random approach (select VPs randomly). 

Both approaches are highly efficient at acheiving high precision. When using a greedy approach, one VP provides the same result as using all VPs. The random approach, which requires no a priori knowledge, can predict 87.41\% of routers to a single country (compared to the optimal \pctPredSingleCountry in Fig. \ref{fig:results_iterative_analysis_stats_country}). Further, at most 6 randomly selected VPs are as good as using all of them. 

By comparison, Wang~\etal~\cite{wang2011towards} and other geolocation approaches that use active measurements~\cite{eriksson:posit,wong:octant} perform measurements from \emph{all} available VPs (see Eriksson~\etal~\cite{eriksson:posit}, which evaluates such approaches using a minimum of 25 VPs). Wong~\etal~~\cite{wong:octant} found that Octant can locate 80\% of target addresses using only 10 VPs, but this is still an order of magnitude larger than our approach. In short, \sysname provides high accuracy without needing a large distributed set of vantage points or large numbers of measurements, thus reducing the barrier to deployment.

\begin{table}[t]
\tabletextsize
\setlength\tabcolsep{4pt}
	\centering
	\small
	\rowcolors{2}{gray!10}{white}
	\begin{tabular}{ p{1.5cm} | p{1.5cm} | p{1.5cm} | p{1.7cm}}
		%\hline
        \textbf{Source}   & \textbf{SoL Viol. (\%)} & \textbf{Inconsistent IPs (\%)}   & \textbf{Affected Paths (\%)} \\     \hline
        EdgeScape         & 0.1      & 7.6        & 37.4        \\ 
        IP2Location       & 1.4      & 7.3        & 31.2        \\ 
        DB-IP             & 6.4      & 14.7        & 60.5        \\
        Maxmind GeoIP2*   & 10.0     & 21.2        & 64.3        \\
        Maxmind GeoLite2  & 13.5     & 21.1        & 64.1        \\
        IP Info	          & 14.0     & 21.5        & 65.6        \\
		IPligence*        & 13.2     & 23.1        & 66.7    \\
%        \textbf{Source}   & \textbf{SoL Viol. (\%)} & \textbf{Consistent (\%)}   & \textbf{Affected Paths (\%)} \\     \hline
%        EdgeScape         & 0.1      & 92.4        & 37.4        \\ 
%        IP2Location       & 1.4      & 92.7        & 31.2        \\ 
%        DB-IP             & 6.4      & 85.3        & 60.5        \\
%        Maxmind GeoIP2*   & 10.0     & 78.8        & 64.3        \\
%        Maxmind GeoLite2  & 13.5     & 78.9        & 64.1        \\
%        IP Info	          & 14.0     & 78.5        & 65.6        \\
%		IPligence*        & 13.2     & 76.9        & 66.7    \\
		%\multicolumn{4}{l}{}																					\\
	\end{tabular}
	\caption{\textbf{Comparison with other approaches.} {\sl While many geolocation databases are highly consistent with \sysname, their inconsistencies affect large fractions of paths in our dataset. Further, several databases exhibit SoL violations for a substantial fraction of IPs in our datatset; we know in these cases that the databases are incorrect. * from \alidadeSourcesMonthYear}}
	\vspace{\postfigspace}
	\label{tab:comparison-final}
\end{table}

\section{Case Studies}
\label{sec:analysis}

\begin{table}[tb]
\tabletextsize
\setlength\tabcolsep{4pt}
	\centering
	\small
	\rowcolors{2}{gray!10}{white}
	\begin{tabular}{ p{1.4cm} | l | p{1.6cm} | p{2.5cm}} %{ l | l | l | l}
		%\hline
		\textbf{Source}  & \textbf{Dest.} & \textbf{Interesting Detours} & \textbf{Cases (Detours/ Total Traceroutes)}			\\ 			\hline
		BR	       & RU     & US, FR, DE  & 12 / 12		\\
		PH	       & PH     & HK          & 1 / 12			\\
		CA	       & CA     & US          & 23 / 457	\\
		CZ	       & CZ     & PL          & 24 / 132		\\
		LV	       & FR     & RU          & 2 / 11		\\
		AM	       & BG, LU & RU          & 8 / 20		\\
        CN         & LB, SA & US, UK, FR  & 12 / 12			\\
        CN         & SG, IN & US          & 326 / 326		\\
        CN         & QA     & US, FR      & 8 / 8		\\
  	    SG, JP, CN & PK	    & US, FR      & 22 / 22		\\
		SG         & CN, PH	& US          & 5 / 5			\\
        PH         & PK, LK & US,  UK     & 5 / 5    \\
        PH         & CN, TH & US          & 11 / 11		\\
        PH         & LB     & FR          & 9 / 9		\\
		TW         & TH		& US          & 3 / 3		\\
		GH	       & MW     & UK, FR      & 312 / 312		\\
		GH	       & TN     & UK          & 58 / 58		\\
		ZA	       & RE     & UK, FR      &	17 / 17		\\
        AR, JP     & RE     & US, FR      & 11 / 11 \\
        NZ         & RE     & BR, US, FR  &	7 / 7 \\
		% \hlinepk
	\end{tabular}
	\caption{\textbf{Interesting Cases.}
		{\sl Path starting and ending in Asia/Africa detour through the US and/or EU. Paths starting and ending in the EU detour through Russia. Some Canadian paths detours the US, a Philippines path detours through Hong Kong, and paths starting and ending in Oceania take a circuitous trip around the globe. }}
	\vspace{\postfigspace}
	\label{tab:suspicious_detours}
\end{table}

%\subsection{Geopolitical Case Studies}\label{subsec:analysis-case-studies}

%To inform this analysis, we conducted a traceroute campaign that focused on measuring from our sources to destinations in the same country as sources (to determine if Internet paths nonetheless left the country), destinations hosting a country's national websites (to increase the chance of a traceroute terminating in that country), and destinations that are popular websites in each measured country (to evaluate the countries traversed by a large fraction of web traffic). In total, we performed 250,000 traceroutes over a period of 7 days in November, 2016\drc{Still using this?} from 87 PlanetLab vantage points in 20 countries to approximately 170 destination countries.
%\drc{Is this a third dataset? Or is it the same as 5.3?} \muz{This is a superset of that dataset but it was done 1 week after that. This not only contains unique Alexa websites but also the some federal government websites(with location validated)}

%\drc{Condense all ``interesting'' cases into a single table or list, so we only have to use the word ``interesting'' once.}

We now use \sysname and the dataset of
\S\ref{subsec:constraint-refinement}, to study paths with interesting properties. We focus on \emph{detours}---paths that traverse at least one country that is not the source or destination country. Paths that start and end in the same country (\eg Canada) but detour through another (\eg the US) are especially interesting, because such ``purely domestic traffic'' is subject to the surveillance and censorship regime in the detoured country. We are also interested in paths that transit multiple continents.   Table~\ref{tab:suspicious_detours} lists sources, destinations, detour countries for several paths in our dataset. Each path described in Table~\ref{tab:suspicious_detours} has been manually validated using hostnames, pings from multiple vantage points, and review of RTTs between hops and next-hop location. We discuss the a few of these paths in more detail below.

%First, we considered pairs of countries with poor or no diplomatic relations at the time of writing. In
%Second, if the source and destination are in the same country, but the path leaves the country, we mark it as ``Country'' in Table~\ref{tab:suspicious_detours}  while country-level detours seen only in reverse traceroutes\cite{revtr} are marked as ``Country-Revtr''. We likewise flag geographic detours that leave a continent before returning to the same continent as ``Continent'' and paths transiting between \emph{two} continents through \emph{multiple} continents before reaching their destination as ``Circuitous''.
%\drc{What others?} \muz{1) Different continent 2)I'm not sure if we can say that any of the intelligence agencies in countries X,Y,Z are powerful and could possibly manipulate traffic, where the countries. X,Y, Z is a list of a) Nuclear powers, or b) highest spenders on defence as a percentage of gdp}

\para{BRICS.} Around late 2014, the BRICS countries (Brazil, Russia, India, China, and South Africa) were reportedly planning to build a undersea fiber cable that would interconnect them, while avoiding the US and Europe~\cite{BRICScable}.  Thus far, however, this effort has come to nought. In our dataset, all the traceroutes from Brazil to Russia transited through the United States and France. We also saw US detours in all paths from China to India.

\para{Russia.}   Paths that start and end in Europe but detour through Russia are interesting, given  reports of potential Russian meddling in European elections~\cite{marconRussia,germanyElectionRussia}. That said, Russia is home to one of the world's largest IXPs (MSK-IX)~\cite{mskIX}, which could explain why we saw several detours from within the EU to Russia.  Some of the paths between Latvia and France are carried by a Russian Telecom company through Russia. In addition, a set of the paths between Armenia and Luxembourg or Bulgaria travel through Russia.

\para{Asia to the US.}  We identified several cases where traffic originating and terminating in different Asian countries detours to the US.  In fact, we found detours to the US in \emph{all} the traceroutes in our dataset from China to its neighbors (India, Pakistan, and Philippines) and from China to the Middle East (Saudi Arabia, Lebanon, and Qatar). This is significant because foreign communications that transit the US are subject to warrantless surveillance under Section 702 of the Foreign Intelligence Surveillance Act (FISA)~\cite{FISAs702}.

\para{Canada to the US. } We identified a path that started and ended in Canada, and took a detour through the US, corroborating evidence provided by the IXmaps project~\cite{ixmaps}.   The US intelligence community recognizes Canadian citizens as ``second party" persons, and thus requires some additional approval (beyond the standard FISA Section 702 authorizations) before they can be ``targeted'' for surveillance~\cite{wapoFlowchart}.

\para{Phillipines to Hong Kong on the reverse path. }  One particularly interesting path started and ended in the Philippines; while its forward path stayed inside the Phillipines, its reverse path passed through Hong Kong.  This path highlights the importance of using reverse traceroute~\cite{revtr} to measure international detours.

\para{Long paths. } Some of the longest (geographically) paths we observed were between Oceania and Africa. For example, we found a path originating in New Zealand that traversed Brazil, the US, France, and then eventually arriving at the destination in Reunion Island. We also found detours through the US, UK, and France for traffic originating and terminating in \emph{different} African countries. The inflated paths between African countries was initially reported by Gupta \etal~\cite{gupta:africaixps}, and tends to result from poor connectivity at regional IXPs.

\para{US traffic transiting abroad.} Of particular interest are cases where traffic starting and ending the US transits through a foreign country, because US surveillance law applies fewer restraints when American's Internet traffic intercepted abroad, rather than inside the US~\cite{MTTLR}.  Our dataset, however, suggests that such cases are rare. We did find some cases (7 out of 32,609 traceroutes) where a traceroute might have exited and entered the United States via a Level 3 router in Toronto. However, the router labeled as Toronto was unresponsive to probes upon validation and so we cannot confirm this finding.

\section{Discussion}
\label{sec:discussion}
%We now briefly discuss caveats and open questions in our work. 

\para{Generalizability.} We used a large set of traceroute measurements to inform the design, implementation, and evaluation 
of our system. However, our results apply necessarily only to the data that we collected. We believe 
that our results will be favorable for other datasets using similar vantage points; however, we cannot make claims about how well 
it will perform in networks and countries that were not in our dataset. Our classifiers were limited by the ground truth labels 
made available to them, and we expect that additional labels will improve our results.

\para{Caveats for machine learning.} We used relatively simple machine-learning classifiers in large part due to their already-high 
accuracy, but it is possible that more advanced techniques would improve the system. An advantage to our decision-tree-based 
approach is that one can inspect the trees to determine whether the classifier is learning something intrinsic to data sources. Such 
analysis can provide confidence in the ability to perform well when provided with different labeled data. 

\para{Improving datasets.} We focused on country-level router geolocation for unidirectional paths based on single snapshots of 
 paths between source/destination pairs. 
% In ongoing work, we are using Reverse Traceroute~\cite{revtr} to investigate the 
% geopolitical implications of bidirectional paths.\drc{I think we have this now...} In addition, 
 As part of ongoing work, we are investigating how geopolitical properties of paths change 
 over time. We will  
 incorporate real-time BGP feeds to investigate suspicious transient geographic detours. We will also expand the set of 
 measured paths to provide greater coverage. Finally, we will investigate how to include crowdsourced data from our 
 online \sysname tool. 
 
\para{Implications.} We highlighted a number of scenarios of ``interesting'' paths in terms of the countries they traversed 
and whether they were inflated. In many of these cases, it is possible that this behavior is normal and even intentional. 
While we tried to highlight cases that we thought were poignant, in general we leave such decisions to individuals with 
sufficient knowledge to draw strong conclusions about implications. 

Certain stakeholders may wish to avoid undesirable paths. We will investigate how to use the PEERING testbed~\cite{peering} to make BGP announcements that 
cause routes to avoid certain countries.  
%* Limitations of ML.
%
%* Reason for decision tree: interpretable, we can look at the data. Our methodology is not biased by the dataset but rather something that can be interpreted.
%
%* We also looked at the limitations of ensemble methods(random forests).
%
%* Limitations of data sources
%
%* Unobserved paths, transient changes
%
%* Legitimate reasons for geopolitical paths(comcast is a giant ISP, US paths)
%
%* Usability of alternative routes
%
%* Limited duration of analysis period

\section{Conclusion}
\label{sec:conclusion}
This paper showed that one can reliably predict the countries visited by routers along a traceroute path using collection of unreliable geolocation sources, when paired with speed-of-light constraints and machine learning. We designed and built a system, \sysname, that does this, and demonstrated that it is accurate, precise, and efficient enough to provide information for submitted traceroutes interactively. We showed that its accuracy is substantially better than standard geolocation sources. We also used our system to evaluate the implications of the geopolitical paths our system identifies, revealing potential security, privacy, and performance issues. 
 %As part of future work we are incorporating larger sets of paths and investigating how their geography changes over time.   

\section{Acknowledgments}
\label{sec:acknowledgements}
This work is partially supported by NSF awards CNS-1618955 and CNS-1405871. We thank the anonymous reviewers for their feedback. We also thank IPInfo\cite{ipinfo} for providing us with their geolocation API.

%{\bibliographystyle{acm}
%\bibliography{choffnes-all}}
\citestyle{acmnumeric}
\bibliographystyle{ACM-Reference-Format}
\bibliography{choffnes-all}

%\cleardoublepage
\clearpage
\begin{appendices}
%\part{\\APPENDIX}
%\part*{APPENDIX}
\section*{\\APPENDIX}
\label{sec:appendix}

\section{OpenIPMap Analysis}
\label{sec:appendix:open-ip-map-eval}

	We evaluated the accuracy of OpenIPMap~\cite{openipmap} before considering it as a source of reliable data for training our classifier, because this data source uses crowdsourced labels that are not independently validated (outside of this study). As a first step in the analysis, we removed the IP addresses with wrongly labeled countries from OpenIPMap that violated SoL constraints. Then we performed a two-fold analysis to evaluate the accuracy of classifiers with and without OpenIPMap. Our results show that using filtered OpenIPMap data when training a Decision Tree classifier improves accuracy, even if a significant fraction of labels are wrong.
	
\subsection{Data Sources}
\label{subsec:appendix:open-ip-map-eval:dataset}

	We use the two data sources as described in Section~\ref{sec:dataset}. These data sources are as follows.
	
	\para{Reliable Sources/Manually Labeled.} These are the IP addresses that we manually labelled using traceroutes and ping measurements, hostname entries, and IP prefixes to responsive routers, as shown in Table~\ref{tab:ground_truth}. This includes IXP locations. 
	
	\para{OpenIPMap.} These are the IP addresses that were collected from OpenIPMap~\cite{openipmap} and validated using SoL constraints, as shown in Table~\ref{tab:ground_truth}.

	\para{Entire Dataset.} This dataset is a \textit{union} of our \textit{reliable data} and \textit{OpenIPMap} datasets, as shown in Table~\ref{tab:ground_truth}.

\subsection{Evaluation}
\label{subsec:appendix:open-ip-map-eval:evaluation}

	To evaluate the effect of using OpenIPMap~\cite{openipmap} as our ground truth, we looked at two aspects of the dataset; poisoning the reliable data, and evaluating the accuracy of this reliably labeled dataset using OpenIPMap. Our overall results show that using OpenIPMap as training data has a net effect of improving accuracy and coverage compared to not using it.

\subsubsection{Poisoning}
\label{subsec:appendix:open-ip-map-eval:evaluation:poisoning}
	To understand the effect of incorrect labels provided by OpenIPMap, we evaluate the resilience of our classifier and its ability to correctly predict the true country despite given wrong training labels.
	
	We take our manually labeled dataset (reliable data) and ``poison'' different percentages of the data by randomly assigning an incorrect label.   Then use 10-fold cross validation on each corresponding Decision Tree classifier to identify how the poisoned data impacts accuracy.
	
	We evaluate four categories of predictions; i) where classifier was able to predict the correct country despite being provided with a wrong label, ii) where classifier predicted to the poisoned country (for the routers where it was poisoned and the true label where it was not poisoned), iii) where the predicted country was neither the true country nor the poisoned country and, iv) where classifier was unable to predict any country and suggested an "unknown country" label.
	
\begin{figure}[tb]
	\centering
	\includegraphics[width=1.05\columnwidth]{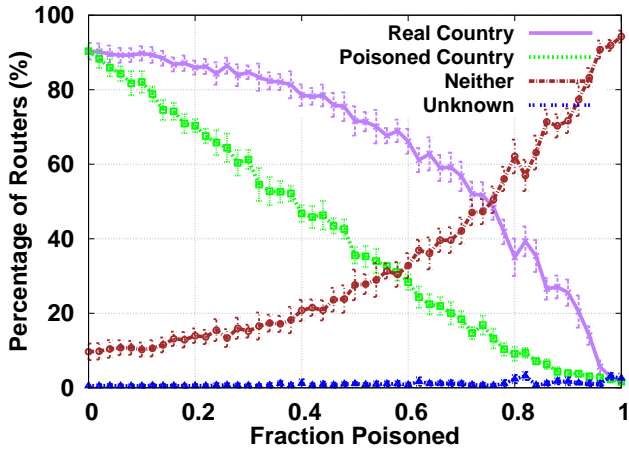}
	\caption{\textbf{Poisoning the dataset} 
		{\sl When 10\% of reliable dataset is poisoned only, the overall accuracy decrease is only 0.6\% (less than half of a standard deviation) showing that our classifiers are resilient to poisoning and can filter out incorrect labels during the training phase.}}
	%\vspace{\postfigspace}
	\label{fig:results_poison_ground_accuracy}
\end{figure}

	Figure~\ref{fig:results_poison_ground_accuracy} shows the effect of poisoning on the reliable data for all four categories mentioned. With 10\% of the labels being incorrect, the accuracy decreased by 0.6\% while with 20\% the data being poisoned, the accuracy decreases by 4.1\%. Even when the 40\% of the dataset is poisoned, the accuracy of the classifier remains well above 80\% showing that our classifiers are resilient to reasonably low levels of incorrect labels.

\subsubsection{Accuracy Analysis for Training on Datasets}
\label{subsec:appendix:open-ip-map-eval:evaluation:training}

	To further understand the affect of OpenIPMap on the accuracy, we evaluated the accuracy of the classifier with and without OpenIPMap. We evaluate the accuracy of the classifiers by training the classifiers using reliable data, OpenIPMap, and the aggregation of both OpenIPMap and manually labeled dataset. We then evaluate them and test the accuracy of the prediction for the reliable data.  

	We used 10-fold cross validation (where  \textit{training} and \textit{testing} datasets were the same) with a Decision Tree classifier to evaluate the accuracy. 

	Table~\ref{tab:accuracy-train-test-dataset} shows that when our reliable dataset is \textit{tested} on the all the other datasets, the difference in accuracy due to different training dataset is low. The second row shows that when using both OpenIPMap and reliable labels, the accuracy difference compared to using only reliable data (first row) is statistically small---within a standard deviation. We also investigate other combinations of training and testing datasets (last three rows) for completeness. 
	
%	 The accuracy decrease when trained on the entire dataset (with a 10-fold cross validation) as compared to manually labeled (with cross validation) is 1.45\%, which is less than 1 standard deviation in the accuracy. Similarly the accuracy decreases from 88.91\% to 83.39\% when trained on OpenIPMap. This is because both are relatively independent datasets (Table \ref{tab:ground_truth}). This decrease is still insignificant when compared to accuracy increase \sysname provides over geolocation sources.

\begin{figure}[tb]
	\centering
	\includegraphics[width=1.05\columnwidth]{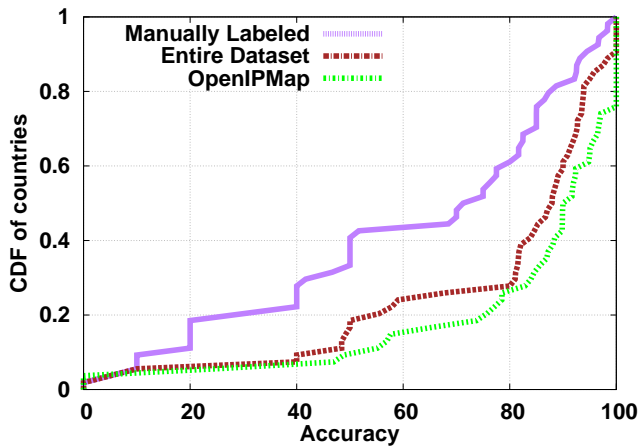}
	\caption{\textbf{Country-level accuracy of the classifier} 
		{\sl when it was trained on the reliable (manually labeled) dataset, OpenIPMap, and the entire dataset (OpenIPMap and manually labeled) and then tested using the manually labeled only. OpenIPMap has the best country-level accuracy while using manually labeled data alone decreases the per-country accuracy for the dataset.}}
	%\vspace{\postfigspace}
	\label{fig:results_dataset_comparison_accuracy_cdf_test_ground_truth}
\end{figure}

	Not only is the difference in accuracy low when including OpenIPMap, the resulting coverage in terms of per-country accuracy is \emph{substantially better}. Figure~\ref{fig:results_dataset_comparison_accuracy_cdf_test_ground_truth} shows the CDF of the country-level accuracy when we \textit{tested} the accuracy of reliable dataset by \textit{training} the classifier on reliable data, OpenIPMap, and the entire dataset. OpenIPMap, despite having the lowest overall accuracy in Table~\ref{tab:accuracy-train-test-dataset} outperforms other training datasets (curve to the bottom right), closely followed by the entire dataset. When trained on OpenIPMap and entire dataset, the classifier was able to predict 73\% and 72\% countries with at least 80\% accuracy, respectively, while this accuracy level is achieved only for 39\% of the countries when the classifier was trained on the reliable data.

	The Table~\ref{tab:accuracy-train-test-dataset} and Figure~\ref{fig:results_dataset_comparison_accuracy_cdf_test_ground_truth} show that using the \textit{entire dataset} rather than \textit{reliable data} alone provides higher \textit{per-country} accuracy with an insignificant decrease in the \textit{overall} accuracy.

% \muz{Remove this paragraph?} 
 We also investigated the accuracy of OpenIPMap when trained on our reliable data, but omit the results because a significant number of countries in the OpenIPMap dataset are not included in the training dataset.

\begin{table}[tb]
	\centering
    \rowcolors{2}{gray!10}{white}
	\begin{tabular}{l | l | l}
		\textbf{Training Data}    & \textbf{Testing Data}     & \textbf{Accuracy (\%)}     \\ \hline
		Reliable Data & Reliable Data & $88.91 \pm 1.56$ \\
		Entire Dataset   & Reliable Data & $87.46 \pm 0.93$ \\
		OpenIPMap        & Reliable Data & $83.39 \pm 0.99$ \\
		Entire Dataset   & Entire Dataset   & $88.45 \pm 1.01$ \\
		OpenIPMap        & OpenIPMap        & $87.32 \pm 0.69$ \\
%		Manual Label & OpenIPMap        & $83.75 \pm 0.29$
	\end{tabular}
	\caption{\textbf{Accuracy for different training-testing datasets} 
		{\sl with 10-fold cross validation. The decrease in accuracy is relatively insignificant when the we compare the accuracy of testing the manually labeled data set trained on manually labeled data  against training on the entire dataset.}}
	\label{tab:accuracy-train-test-dataset}
\end{table}

%%%%%%%%%%%%%%%%%%%%%%%%%%%%%%%%%%%%%%%%%%%%%%%%%%%%%%%%%%%%
%%%%%%%%%%%%%%%%%%%%%%%%%%%%%%%%%%%%%%%%%%%%%%%%%%%%%%%%%%%%
%%%%%%%%%%%%%%%%%%%%%%%%%%%%%%%%%%%%%%%%%%%%%%%%%%%%%%%%%%%%

\begin{comment}
\section{}
\label{sec:appendix:temp}

	Figure~\ref{fig:replacement:results_iterative_analysis_stats_country} to be used as a replacement figure for Figure~\ref{fig:results_iterative_analysis_stats_country}

\begin{figure}[]
	\centering
	\includegraphics[width=1.1\columnwidth]{results_iterative_analysis_stats_country_overall_second}
	\caption{\textbf{Impact of using speed of light (SoL) constaints}. 
		{\sl By incorporating SoL constraints to rule out infeasible router locations, then using this information to retrain our classifiers, our system predicts exactly one country \pctPredSingleCountry of the time, compared to the initial classifier that does so only 63.7\% of the time.  \muz{added} This is the figure to replace \ref{fig:replacement:results_iterative_analysis_stats_country} if needed}}
	\vspace{\postfigspace}
	\label{fig:replacement:results_iterative_analysis_stats_country}
\end{figure}
\end{comment}

%%%%%%%%%%%%%%%%%%%%%%%%%%%%%%%%%%%%%%%%%%%%%%%%%%%%%%%%%%%%
%%%%%%%%%%%%%%%%%%%%%%%%%%%%%%%%%%%%%%%%%%%%%%%%%%%%%%%%%%%%
%%%%%%%%%%%%%%%%%%%%%%%%%%%%%%%%%%%%%%%%%%%%%%%%%%%%%%%%%%%%

\begin{table}[tb]
    \centering
    \rowcolors{2}{gray!10}{white}
    \begin{tabular}{l | l}
        \textbf{Feature}    & \textbf{Importance (\%)}      \\ \hline
        IP2Location Country & $71.00$ \\
        IP address          & $6.70$ \\
        DDec  Country       & $6.44$ \\
        DB-IP Country       & $5.00$ \\
        ISP Name            & $4.26$ \\
        IPInfo  Country     & $1.83$ \\
        AS Country          & $0.90$ \\
        IP Prefix           & $0.76$ \\
        AS Name             & $0.83$ \\
        ISP Country         & $0.46$ \\
        EurekAPI            & $0.42$ \\
        Maxmind GeoLite     & $0.41$ \\
        AS Number           & $0.23$ \\
        AS Registry         & $0.22$ \\
        ISP size (\# ASes)  & $0.17$ \\
        ISP customer cone   & $0.17$ \\
        ISP city            & $0.09$ \\
        ISP Region          & $0.08$
    \end{tabular}
	\caption{\textbf{Feature importance for all features} {\sl in the \emph{default} classifier (with all labeled data) in the \ensemble. The country predicted by IP2Location is the most important feature in the classifier predictions.}}
    \label{tab:feature-importance-all-default}
\end{table}

\section{Feature Importance}
\label{sec:appendix:feature-importance}

	In a Decision Tree classifier, different features have different weights associated with them. One feature is, sometimes, preferred over another to make the decision. 
	
	To understand the effect of all the features on the classifiers, and identify the most important ones in deciding the country label by our classifier, we perform feature selection analysis on the \emph{default} classifier.
	
	We train the classifier on our ground truth training dataset and then analyze the assigned weights. We do not enforce any weights and let the classifier decide the power of each feature.
	
	Table~\ref{tab:feature-importance-all-default} summarizes the importance of each feature for the \emph{default} (all labeled data) classifier. Our results show that IP2Location is the primary feature used by the classifier decide a country label, with IP2Location being the primary decision factor for 71\% of the rules in the classifier. The IP address and DDec~\cite{ddec} each contribute to 6.7\% and and 6.4\%, respectively. The reason for low contribution by DDec is because the hostnames are available for only 7\% of routers in our dataset.
	The feature importance for \emph{balanced} with maximum accuracy sampling approach and  \emph{unbalanced} with knee-based sampling approach was similar but not identical to the \emph{default} classifier, whereas, for the classifier trained with \emph{equal number of training instances per country}, the percentage was less skewed (in favor of any specific source) giving higher weights to ISP Country and ISP size.

%%%%%%%%%%%%%%%%%%%%%%%%%%%%%%%%%%%%%%%%%%%%%%%%%%%%%%%%%%%%
%%%%%%%%%%%%%%%%%%%%%%%%%%%%%%%%%%%%%%%%%%%%%%%%%%%%%%%%%%%%
%%%%%%%%%%%%%%%%%%%%%%%%%%%%%%%%%%%%%%%%%%%%%%%%%%%%%%%%%%%%

\begin{figure*}[!htb]
\minipage[t]{0.32\textwidth}

	\includegraphics[width=1.05\columnwidth]{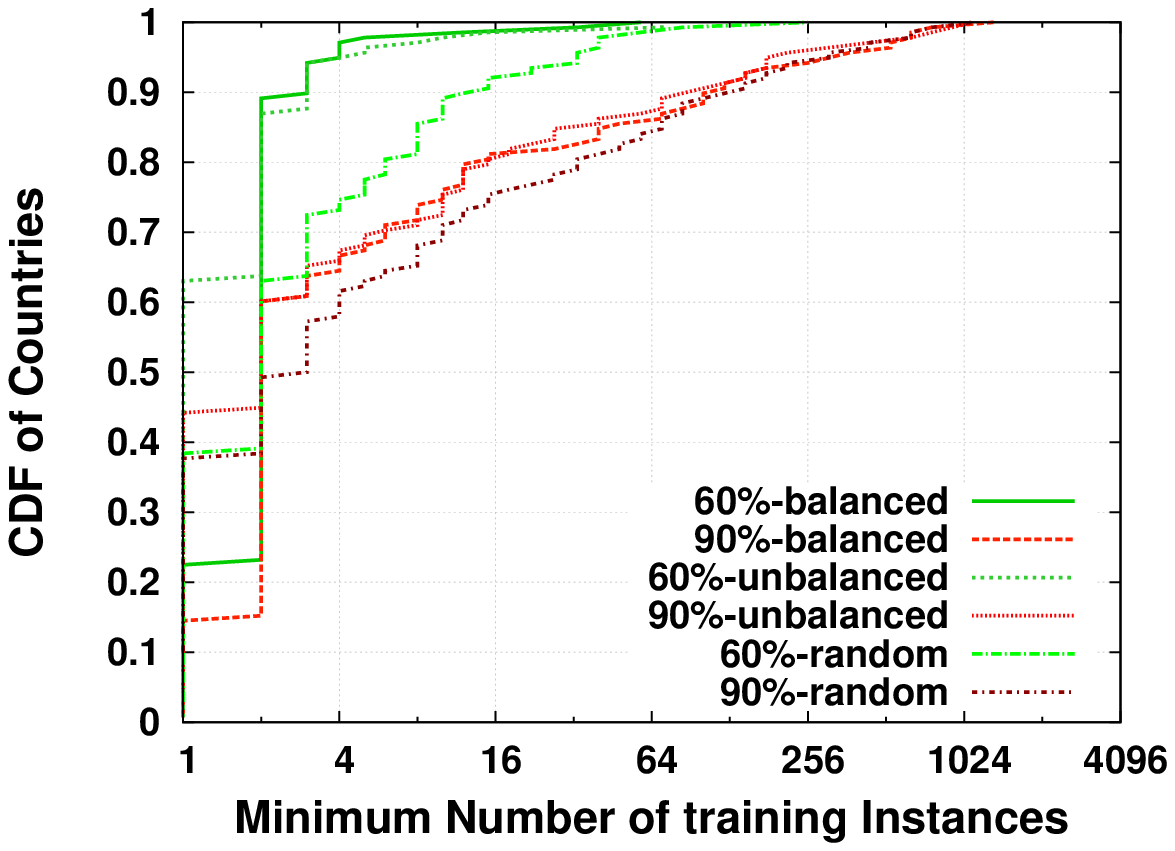}
	\caption{\textbf{Decision tree sensitivity to number of training instances}. 
		{\sl More than two thirds of countries require a small number of instances to achieve 90\% accuracy, but some require 100s of instances to achieve high accuracy.}}
	\vspace{\postfigspace}
	\label{fig:results_classifier_sensitivity_training_1}
\endminipage\hfill
\minipage[t]{0.32\textwidth}
	\includegraphics[width=1.05\columnwidth]{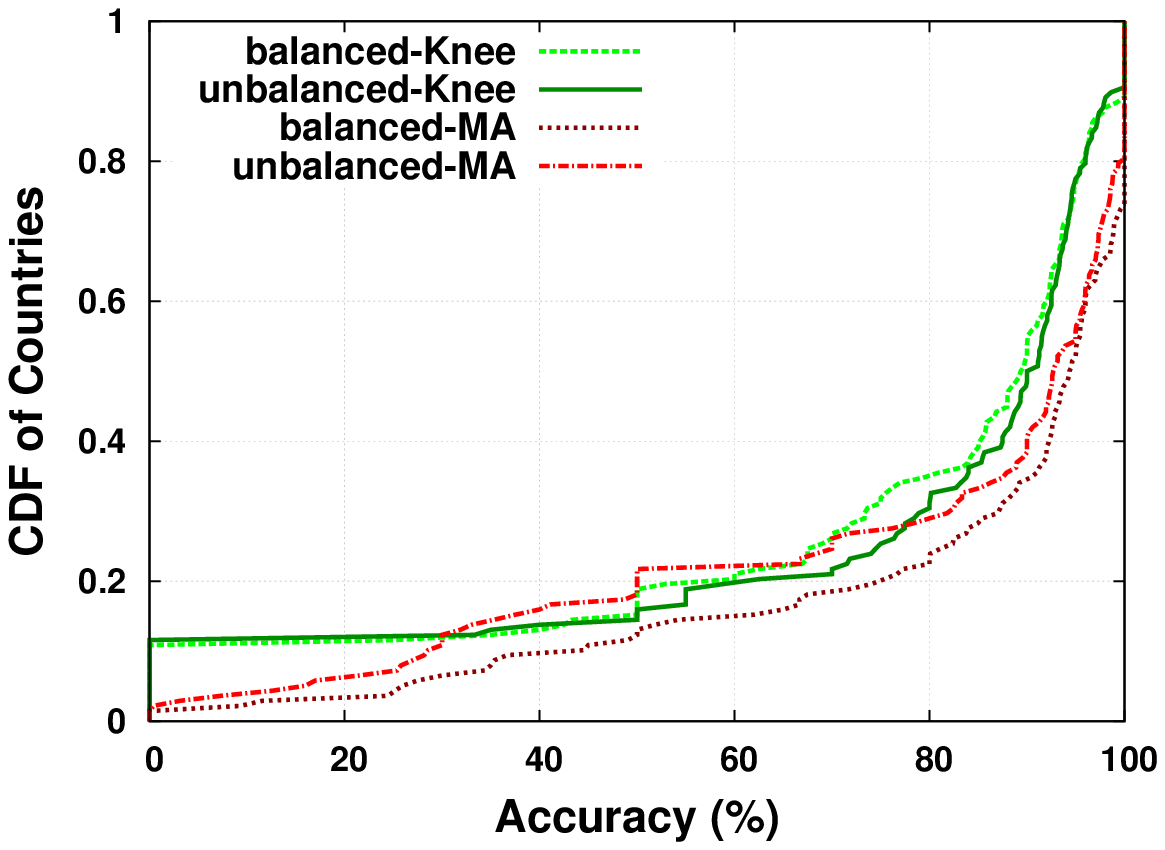}
	\caption{\textbf{Decision tree sensitivity to algorithm for selecting training instances}. 
		{\sl Comparing the accuracy of the countries under different selection methods for the training dataset. }}
	\vspace{\postfigspace}
	\label{fig:results_classifier_sensitivity_training_2}
\endminipage\hfill
\minipage[t]{0.32\textwidth}%
	\includegraphics[width=1.05\columnwidth]{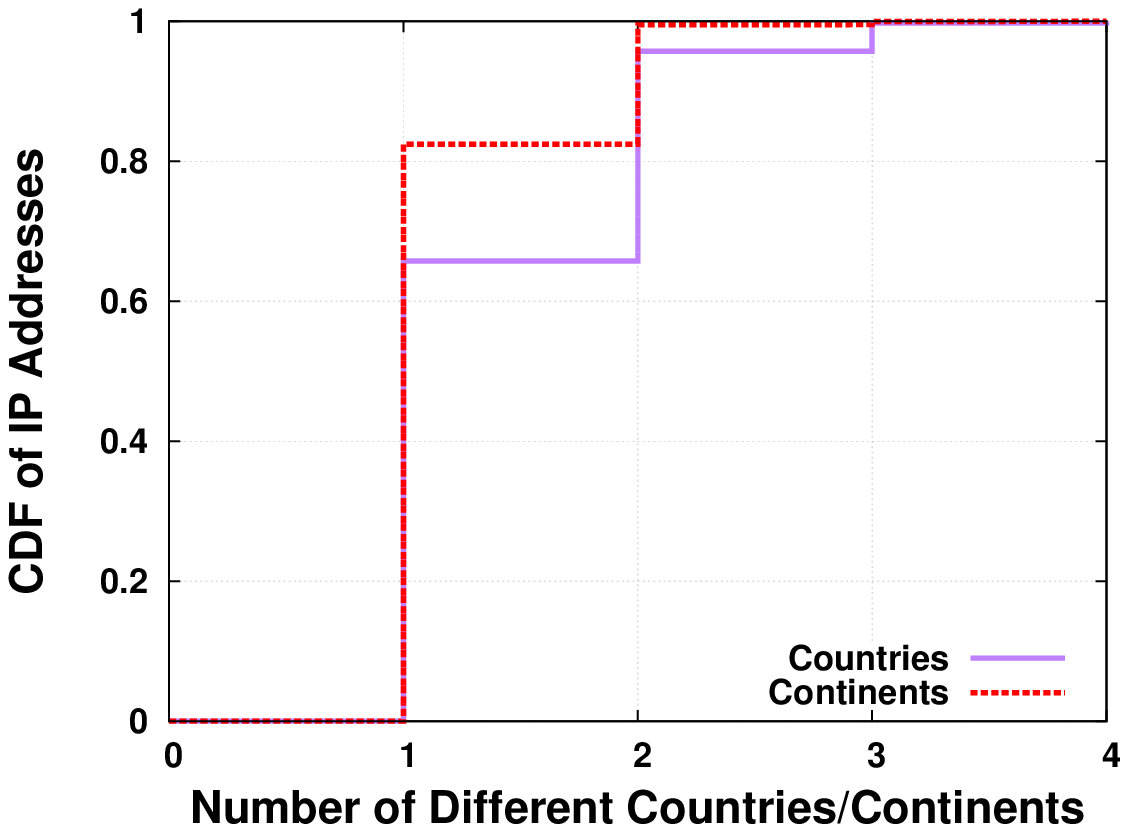}
	\caption{\textbf{Precision}. 
		{\sl \sysname is precise in that the total number of countries and continents mapped by the \ensemble classifier for each unique IP address tends to be one most of the time.}}
	\label{fig:results_classifier_precision}
\endminipage
\end{figure*}

\section{Ensemble Classifier}
\label{sec:appendix:ensemble}

This section provides supplementary details of decisions behind the  construction of \ensemble classifier, details on the sensitivity analysis of the classifiers, the reasoning behind the design decisions involved in the construction of the \ensemble and the evaluation of the precision of the \ensemble.

%All the classifier related decisions addressed here have been aggregated together. All the subsections of this appendix, however, are independent of each other.

\subsection{Classifier Selection}
\label{sec:appendix:classifier-selection}
A simple statistic of the \emph{overall} classifier accuracy in Table~\ref{tab:result_classifier_accuracy} is potentially biased by the number of samples per country in our dataset and is not necessarily instructive of whether the classifier offers high accuracy across a wide range of countries. 

To investigate this, we compare the top-performing classifiers from Table~\ref{tab:result_classifier_accuracy} by training them with data by  \textit{randomly sampling} training data from our dataset.\footnote{Note that we  evaluate additional training dataset selection approaches to understand classifier sensitivity in Section~\ref{sec:evaluation}.} We then find the accuracy of each classifier according to the fraction of a country's routers that are predicted correctly. We plot this in Figure \ref{fig:results_classifier_comparison_accuracy_all} as a CDF of accuracy per country, where the x-axis represents accuracy and y-axis is the fraction of countries with router locations predicted correctly at least $x$\% of the time.

For $x=0.9$, the figure shows that Decision Tree and Random Forest achieves an accuracy of 90\% or better for 35\% of the countries. The 1-Nearest Neighbor algorithm achieves the same level of accuracy for only 3\% of the countries and AdaBoost never achieves 90\% accuracy. 

Figure \ref{fig:results_classifier_comparison_accuracy_all} shows that all the classifiers fail to predict 23\% of the countries.  In our dataset, these countries are the ones that have only one or two training instances. 

To study the impact of bias, we plot Figure~\ref{fig:results_classifier_comparison_accuracy}, with the axes similar to  Figure~\ref{fig:results_classifier_comparison_accuracy_all}, as a CDF of accuracy per country by using an \emph{equal number of training examples per country} (\ie by oversampling from countries with few routers and undersampling from those with large numbers of routers). 
%where the axes are similar to  Figure~\ref{fig:results_classifier_comparison_accuracy_all}. The figure shows that 
Decision Trees and Random Forests still outperform other classifiers, achieving an accuracy of 90\% or better for 76\% of the countries. However, we also found that accuracy decreased for certain countries compared to the \emph{default} classifier, thus a single classifier does not necessarily perform well for \textit{all} countries. 

%\drc{Did you actually check for this?} \muz{yes, IP2Location.}

%These figures show that a training data bias affects the per-country accuracy for different countries and 
%single one-size-fits-all classifier doesn't capture the broad scope of instances that are possible.\drc{I can't quite follow the second point, nor do I think you have shown it here.}
%\muz{added ends}

\subsection{Microbenchmarks and Classifier Sensitivity}
\label{sec:appendix:results_classifier_training_sensitivity}

We investigate the effectiveness of the individual components of our \ensemble classifier and their sensitivity to the training dataset under different data \emph{sampling} techniques. 
We focus on the empirically derived number of training instances \emph{for a country} that provide maximum per-country accuracy (\emph{maximum accuracy}) and diminishing returns for increasing number of training instances (\emph{knee}) approach from Section~\ref{subsubsec:methodology-classifier}.

First, we investigate the number of instances required to achieve a per-country accuracy of 60\%  and 90\%. The outcome depends on how we allocate training instances to other countries, so we investigate three schemes: using the same number of samples as in the target country (\emph{balanced}), double the instances in other countries (\emph{unbalanced}), and a random number of samples (\emph{random}).  

Figure \ref{fig:results_classifier_sensitivity_training_1} shows the minimum number of training instances required to achieve a given per-country accuracy level (x-axis), as a CDF over all countries in our dataset (y-axis). Unsurprisingly, lower accuracy thresholds require fewer training instances. Focusing on the 90\% accuracy threshold, about two thirds of countries need eight or fewer training instances. This is encouraging because providing ground truth labels is a time-consuming, manual process for many router IPs. Note, however, that there is a long tail to the graph, indicating that some countries, like United States, France, Germany and Russia, need large numbers (hundreds or thousands) of instances to achieve high accuracy. 
%Interestingly, the choice of selecting training instances for other countries does not have much impact on accuracy. %\drc{Does this mean we can cut a couple of paragraphs and figures and just summarize the results?} 

We now evaluate our strategies for automatically selecting the best instances to use for training. Recall from  Section~\ref{subsubsec:methodology-classifier} that we incorporate two schemes: maximizing per-country accuracy (\emph{maximum accuracy}) and finding the point of diminishing returns for accuracy improvement (\emph{knee}). %\drc{Again, we can probably condense this and cut the graph. We don't want to overload the reader.}
 Figure \ref{fig:results_classifier_sensitivity_training_2} shows the per-country accuracy for each approach using a CDF of the per-country accuracy (x-axis) over all countries in our dataset (y-axis). 
%Curves toward the bottom right are more accurate (better). 
Overall, we find that \emph{random} performs systematically worse (not shown), while  \emph{maximum accuracy} and \emph{knee} approaches on both \emph{balanced} and \emph{unbalanced} sampling methods yield similar results.

\subsection{Ensemble Construction}
\label{sec:appendix:ensemble-construction}

In constructing the ensemble, we investigate how to combine the classifiers to optimize between accuracy and precision.

Figure \ref{fig:results_classifier_sensitivity_training_2} shows the approaches to achieve a maximum gain in the accuracy per country, however, it hides the overlap for the predicted countries between different approaches. 
 
We now analyze the impact of adding each of these classifiers along with the classifiers trained using the same number of instances, to a classifier trained on the entire training dataset (\ie the \emph{default} classifier).  

We plot a figure similar to Figure~\ref{fig:results_iterative_analysis_stats_country} to evaluate this (not shown). Our study found that at least $6$ classifiers with \emph{equal number of training instances for each country} are required to achieve convergence in the marginal increase in overall accuracy, while keeping the average number of countries predicted to a minimum, as more classifiers are added to the \emph{default} classifier. These classifiers are trained using \emph{mean} of the \emph{number of training instances per country} in the dataset, since for our dataset, the mean is a better representative to have a significant number of training instances as compared to the median or the mod.

We evaluate candidates for the ensemble by combining i) $6$ classifiers trained using different subsets \emph{with equal number of instances per country}, ii) all $4$ classifiers from Figure \ref{fig:results_classifier_sensitivity_training_2} and, iii) \emph{default} classifier. This gave us a total of $11$ classifiers.

To find the the minimum number of classifiers to achieve convergence in the increase in accuracy, we make \emph{n}-length combinations of using all the sets of classifiers (all $11$ classifiers).

As seen in Figure~\ref{fig:runme_optimal_classifiers_delta}, we achieve diminishing returns in the accuracy as accuracy more classifiers are added. At $x = 3$, the increase in average accuracy is roughly $1.1\%$ while it decreases to $0.3\%$
 for $x=4$ and beyond this point the increase  in accuracy is negligible.

We pick the first \numClassifiers classifiers to form our \ensemble classifier. 
It contains \emph{default}, \emph{balanced} dataset bias with \emph{knee} sampling approach, \emph{unbalanced} with \emph{maximum accuracy} approach, and one classifier with equal number of training instances for all countries.

\subsection{Precision}
\label{sec:appendix:results_classifier_precision}

To evaluate the question of precision of our \ensemble classifier, we plot the distinct number of countries and continents returned by the classifier for each router in our dataset, using a CDF in Figure~\ref{fig:results_classifier_precision}. A point in each curve represents the fraction of IP addresses (y-axis) for which at least $x$ countries were predicted (x-axis). 

The graph shows that a single country is predicted for 65.7\% of router IPs, while a single continent is predicted for 82.4\% of router IPs. Thus, the \ensemble alone is accurate but not precise; further, due to the spread in continents predicted, this precision is not sufficient to reason about geopolitical implications of Internet paths.

%%%%%%%%%%%%%%%%%%%%%%%%%%%%%%%%%%%%%%%%%%%%%%%%%%%%%%%%%%%%
%%%%%%%%%%%%%%%%%%%%%%%%%%%%%%%%%%%%%%%%%%%%%%%%%%%%%%%%%%%%
%%%%%%%%%%%%%%%%%%%%%%%%%%%%%%%%%%%%%%%%%%%%%%%%%%%%%%%%%%%%

\begin{table}[tb]
    \centering
    \rowcolors{2}{gray!10}{white}
    \begin{tabular}{l | l}
        \textbf{Source}                 & \textbf{Accuracy (\%)} \\ \hline
        Passport               & $96.47$         \\
        EdgeScape              & $90.93$         \\
%        EdgeScape*             & $88.59$         \\
        IP2Location            & $92.28$         \\
%        IP2Location*           & $88.92$         \\
        DB-IP                  & $86.41$         \\
        Maxmind GeoLite2       & $85.07$         \\
%        Maxmind GeoLite2*      & $80.54$         \\
        Maxmind GeopIP2*       & $81.88$         \\
        IPInfo                 & $84.39$         \\
%        DB-IP*                 & $83.22$         \\
        IPLigence*             & $81.38$         \\
%		\multicolumn{2}{l}{* from \alidadeSourcesMonthYear}	
    \end{tabular}
	\caption{\textbf{Accuracy of \sysname} {\sl as compared to other geolocation sources. \sysname outperforms all other geolocation sources for the labeled data.} * from \alidadeSourcesMonthYear}
    \label{tab:accuracy-comparison-sysname-others}
\end{table}

\section{Accuracy Comparison with Geolocation Sources}
\label{sec:appendix:geolocation-sources-accuracy-comparison}

	This section expands on our analysis at the end of Section~\ref{subsubsec:results-comparision-alternate}. We sampled and labeled the inconsistent locations between \sysname and geolocation services in a manner similar to Section~\ref{sec:dataset}. Of the 80 inconsistent locations between \sysname and EdgeScape, \sysname was correct for 60 of those, while EdgeScape was correct for 18, and both were wrong for 2 of those locations. Similarly, of the 54 inconsistencies with IP2Location labeled across 41 countries, \sysname was correct for 39 of those, IP2Location for 13, and none of them for 2 locations. For Maxmind-GeoLite2 database, of the 78 labels across 54 countries, \sysname successful for 65 and Maxmind for only 10 cases with both being wrong for 3 cases. As for DB-IP, 86 instances of inconsistent locations for 56 countries was labeled, \sysname correctly located 68 cases while DB-IP was successful for only 11 cases. IPInfo showed similar results as DB-IP as 78 out of 97 labeled inconsistencies were correctly predicted by \sysname while IPInfo was correct for only 10 cases.

	When all the inconsistent labels were fed \sysname, it was able learn and predict 3 more consistent cases with EdgeScape, 2 more for IP2Location, DB-IP and Maxmind-GeoLite 2 and 1 more case for IPInfo.

	To further establish a better understanding of \sysname accuracy when compared to other geolocation sources, we manually label 596 IP addresses across 519 ASes and 146 countries (and independent territories) in a manner similar to Section~\ref{sec:dataset}, except we chose 1 IP address per AS instead of 5. We used these labels and found the accuracy of all the geolocation sources including \sysname. 
	
     Table \ref{tab:accuracy-comparison-sysname-others} summarizes the results showing that \sysname had the highest accuracy with 96.47\% IP addresses being labeled correctly, followed by IP2Location at 92.28\% and Edgescape at 90.90\%. 
    
    The results showed that our diversity of probes in different countries allowed us to locate 4.02\% routers correctly where all other geolocation sources failed to predict the correct country. Moreover, for 1.51\% cases, atleast one of the geolocation sources was correct while \sysname failed to predict the correct location.
    
    Of all incorrect by predictions by EdgeScape, it defaulted and provided United States as the predicted country for 3.02\% of the cases (one-third of failed cases). No patterns were found for other geolocation sources.
    
    Looking at the continent-level accuracy, \sysname had highest accuracy for all continents except Europe. EdgeScape had the highest for Europe with EdgeScape predicting 95.35\% routers correctly and \sysname being correct for 93.4\% cases    
        
    %Another interesting thing highlighted was that 71.14\% routers didn't change location between \alidadeSourcesMonthYear and May 2017.

	We analyzed DDec~\cite{ddec} as well. The DDec-interpretable hostname 	information was available for only 8\% of the routers. Of the routers where hostname country was available, \sysname had a consistency of 99.8\% with DDec interpretations.
	
	We intended to include NetAcquity as a comparison point; however, the company would not offer us access to their database at a reasonable price without publication restrictions, which we refused to accept on principle.

%%%%%%%%%%%%%%%%%%%%%%%%%%%%%%%%%%%%%%%%%%%%%%%%%%%%%%%%%%%%
%%%%%%%%%%%%%%%%%%%%%%%%%%%%%%%%%%%%%%%%%%%%%%%%%%%%%%%%%%%%
%%%%%%%%%%%%%%%%%%%%%%%%%%%%%%%%%%%%%%%%%%%%%%%%%%%%%%%%%%%%

\begin{figure}[tb]
	\centering
	\includegraphics[width=1.05\columnwidth]{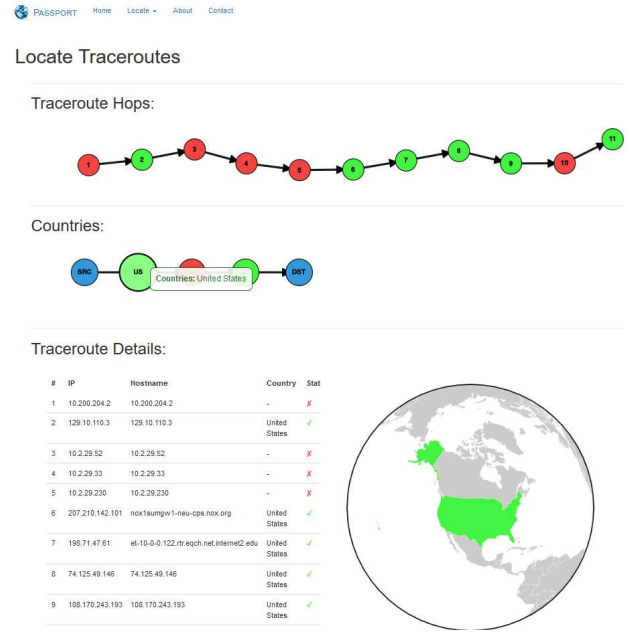}
	\caption{\textbf{Traceroute performed by \sysname.} 
		{\sl The web interface provides a green circle for hops for all the hops that were located, while a yellow one representation ongoing measurement for the router and red one shows private addresses and system failures (with reasoning).
		For countries, the blue circles represent the source and the destination, while the map shows a visualization of the countries traversed in a path.}}
	\vspace{\postfigspace}
	\label{fig:passport_online_system}
\end{figure}

\section{Online System}
\label{sec:appendix:online-system-website}

\sysname system is publicly available for use via a web interface and a REST API. The user interface developer documentation, and the source code, are also public at \url{https://passport.ccs.neu.edu}

	Figure~\ref{fig:passport_online_system} shows a snapshot of the output result from the website of \sysname for a sample traceroute measurement. While the web interface provides a visual representation of the countries visited for a traceroute (or an IP address), the API is more powerful and has the ability to provide predictions individual results by \ensemble, the SoL system, and the overall \sysname prediction.
	
	When a traceroute is provided to website or the API of \sysname, it's parsed for valid router IP addresses and their corresponding RTT measurements, as shown in Figure \ref{fig:online_stage_design}. 
	Using the IP address, the location predictions from geolocation services is collected and this information is used to predict a set of countries by the classifier. 
	The IP address is used to issue new ping measurements from our vantage points. These are combined with the user-provided measurements to construct the SoL constraints for the router location.
	The classifier predictions are then evaluated using these SoL constraints and result is returned.

All this information is also stored for offline analysis and caching is performed at each step to speed up future queries with a flexible cache duration.

Since \sysname incorporates access to some external rate-limited APIs and shared measurement vantage points, new requests in the \sysname core (online system) use a job scheduler.

\end{appendices}

%\theendnotes
\end{document}